\documentclass[11pt,reprint,english,aps,prc,showpacs,letter,longbibliography]{revtex4-2}

\usepackage[     bookmarks,
                 bookmarksopen = true,
                 bookmarksnumbered = true,
                 linktocpage,
                 colorlinks = true,
                 linkcolor = blue,
                 urlcolor  = blue,
                 citecolor = blue,
                 anchorcolor = green,
                 hyperindex = true,
                 hyperfigures]
        {hyperref}

\usepackage{graphicx}
\usepackage{dcolumn}
\usepackage{bm}
\usepackage{float}
\usepackage{amsmath}
\usepackage{booktabs}

\usepackage{color}
\usepackage{graphicx}
\definecolor{dgreen}{cmyk}{1.,0.,1.,0.2}        
\definecolor{orange}{cmyk}{0.,0.353,1.,0.}    

\def \r {\rho}

\newcommand\sect[1]{\section{#1}}

\begin{document}

\title{Sampling nucleon short-range correlations in coordinate and momentum space for Au+Au collision simulations}

\author{Yu-Jing Huang$^{1}$}
\author{Qian-Ru Lin$^{1}$}
\author{Long-Gang Pang$^{1}$}
\email{lgpang@ccnu.edu.cn}

\address{$^{1}$ Key Laboratory of Quark and Lepton Physics (MOE) and Institute of Particle Physics, Central China Normal University, Wu han 430079, China}

\date{\today}%

\begin{abstract}
We develop a nuclear initialization scheme that jointly incorporates spatial nucleon correlations, high-momentum components, predominantly proton–neutron pair assignments, and correlated pair momenta, and implement it in the SMASH transport model. This construction enables controlled comparisons of the effects of different ingredients of short-range correlations (SRCs) on heavy-ion observables. For Au+Au collisions at \(\sqrt{s_{NN}}=2.4\) GeV, SRC initialization increases both charged-pion yields relative to a Woods–Saxon reference with Fermi momenta when the same mean-field parameters are used. The larger relative enhancement of \(\pi^+\) production reduces the \(\pi^-/\pi^+\) ratio. Controlled comparisons identify the modified single-nucleon momentum distribution as the dominant source of the enhancement, with smaller effects from the tested spatial and pair-isospin variations and enhanced production persisting without the imposed pair-momentum structure. Reaction histories link this enhancement to increased baryon-resonance production. 
Stronger high-momentum repulsion counteracts the
SRC-induced pion enhancement by reducing the energy
available in nucleon--nucleon collisions and suppressing resonance production.
The similar enhancement of $K^+$ production with SRC
initialization and its reduction after refitting the
mean-field parameters show that the competing effects
of correlated initialization and mean-field dynamics
extend beyond pion production.

\end{abstract}
\maketitle


\section{Introduction}

Electron-scattering and two-nucleon knockout measurements
reveal a predominance of proton--neutron short-range
correlated pairs with large relative momenta and smaller
center-of-mass momenta in the kinematic regions probed
\cite{Subedi:2008zz,CLAS:2018xvc,CLAS:2018qpc,
Frankfurt:2008zv,Arrington:2022sov,Hen:2014nza,
CLAS:2018yvt,CLAS:2020mom,CLAS:2020rue,Arrington:2011xs}.
The high-momentum components associated with these
short-range correlations (SRCs) are absent from a
zero-temperature local Fermi-gas initialization,
which populates momentum states only up to the local
Fermi momentum \cite{Buss:2011mx,Benhar:1994hw,Rios:2009gb,Hen2017Review}.
In descriptions with realistic nucleon--nucleon
interactions, tensor force and short-range repulsion
generate high-momentum components together with a
depletion of occupation below the Fermi surface
\cite{Ciofi2015Review,Hen2017Review,Schiavilla:2006xx,
Wiringa:2013ala,Lonardoni:2018sqo}.
These effects extend beyond the independent-particle
picture underlying conventional initialization prescriptions
\cite{Benhar:1994hw,Rios:2013zqa,Ryckebusch:2014ann}.
Many-body calculations relate the short-distance two-nucleon
structure to pair distributions at large relative momenta,
providing a connection between the spatial and momentum-space
manifestations of SRCs
\cite{Ciofi2015Review,CiofidegliAtti:1995qe,Alvioli:2016wwp,
Weiss:2015mba,Weiss:2016obx,Cruz-Torres:2019fum,Casale:2023ews}.

In heavy-ion collisions, SRC-related changes in the two-nucleon spatial distribution have been investigated through their effects on initial-state geometry and collective observables~\cite{Alvioli:2011sk,Denicol:2014ywa,Alvioli:2009ab,Broniowski:2010jd,Liu2025IsolatedNN}.
For ultracentral Pb+Pb collisions at $\sqrt{s_{NN}}=2.76$ TeV,
correlated nuclear configurations and minimum-distance
prescriptions modify the initial eccentricities and have been
explored to address the $v_2$--$v_3$ puzzle, namely the difficulty
of reproducing the nearly equal measured elliptic and triangular
flow coefficients~\cite{Denicol:2014ywa,Wang2025Ultracentral}.
Studies of high-energy $^{16}$O+$^{16}$O collisions using nuclear configurations from ab-initio calculations have explored the effects of nucleon correlations on elliptic flow and its correlation with the mean transverse momentum~\cite{Zhang2025AbInitioOO,Liu2026OOSRC}.
Calculations at 5.36 TeV further show that higher-order cumulants
of the initial energy-to-entropy ratio are sensitive to two-nucleon correlations, with this response transmitted to final-state
mean-transverse-momentum cumulants through hydrodynamic evolution
\cite{Li:2025bdn}.
At $\sqrt{s_{NN}}=3$ GeV, SMASH calculations of Au+Au collisions
found relatively small changes in conventional observables,
while a multievent deep-learning analysis identified differences
between correlated and uncorrelated initial configurations
\cite{Huang2025NNCorrelations}.

The high-momentum tail is particularly relevant to particle
production in low- and intermediate-energy nuclear collisions,
especially near production thresholds, where intrinsic nucleon
momenta can appreciably change the energy available in individual
collisions \cite{Yong:2015gma,Wei:2019xdw,Yong:2021nwn}. IBUU calculations of Au+Au collisions at a beam
kinetic energy of 400 MeV per nucleon found that charged-pion
yields and charge ratios are sensitive to the form of the initial
high-momentum distribution, whereas the nucleon flows examined
are comparatively insensitive~\cite{Yang:2018xtl}.
At relativistic energies, free spectator-neutron observables have been proposed to probe the high-momentum fraction and its isospin dependence~\cite{Liu:2026tde}. In UrQMD calculations of Au+Au collisions at a beam kinetic energy of 1.23 GeV per nucleon with in-medium inelastic $NN$ cross sections, adding a high-momentum tail increased both charged-pion yields and lowered their $\pi^-/\pi^+$ ratio~\cite{Li:2026dsx}. This raises the question of how sensitive these yields and their ratio are to the spatial, isospin, and momentum correlations of individual nucleon pairs.

Beyond their effects on collision observables,
SRC-induced changes in the momentum distribution also
modify the kinetic and interaction contributions to the
nuclear equation of state (EOS)~\cite{Cai:2025txx}. In a nonlinear relativistic mean-field model, Cai and Li
found that including SRC-modified momentum distributions
increased the high-density pressure and the maximum
neutron-star mass~\cite{Cai:2015xga}.
Their later study using a non-relativistic Gogny-like
energy-density functional obtained a decrease in the maximum
mass~\cite{Cai:2022grw}.
The latter work attributes the lower maximum mass
in its Gogny-like calculation to stronger softening
of the high-density symmetry energy and a weaker
increase in symmetric-matter pressure than in the
relativistic models~\cite{Cai:2022grw}.
The proton fraction in neutron-star matter and the
crust--core transition density and pressure also decreased,
whereas the radius of a $1.4M_\odot$ neutron star changed little.
These results highlight the interplay between SRC-modified
momentum distributions and the nuclear EOS in different
theoretical descriptions.

The HADES measurements of charged-pion production
and proton flow in Au+Au collisions at
$\sqrt{s_{NN}}=2.4$~GeV provide complementary constraints
on transport descriptions of the same reaction system
\cite{HADES:2020ver,HADES:2022osk}.
A systematic SMASH analysis included a momentum-dependent nucleon
potential, which strengthened the elliptic-flow signal and was
needed to describe the flow data within that framework; the
analysis favored an incompressibility of $K\simeq348$ MeV
\cite{Mohs:2024gyc}.
Beyond constraints from collective flow, pion production
provides complementary information on the momentum
dependence of the mean field~\cite{Hong:2013yva}. 
Using interactions with common bulk nuclear-matter characteristics
but different momentum dependence, LBUU calculations showed that
a potential rising more strongly with momentum
reduces the fraction of high-momentum nucleons during compression,
suppressing $\Delta$ production and the subsequent pion yields
\cite{Li2025MeanFieldPions}.

In this work, we construct nuclear initial configurations that
jointly incorporate spatial nucleon correlations, high-momentum
components, predominantly proton--neutron pair assignments,
and correlated pair momenta, and implement them in the SMASH
transport model~\cite{SMASH:2016zqf}.
We examine how correlated nuclear initialization changes pion
production and the mean-field parametrization required to
describe the data.
We focus on Au+Au collisions at $\sqrt{s_{NN}}=2.4$ GeV,
comparing charged-pion observables and proton flow with
calculations using a Woods--Saxon reference initialization.
Controlled comparisons of selected initialization ingredients
are used to identify the origin of the pion-yield response.
The mean-field parameters are then readjusted using charged-pion
multiplicities and proton elliptic flow, while propagation and
reaction histories are analyzed to investigate changes in
collision energies and resonance production.
Additional calculations at $\sqrt{s_{NN}}=3.2$ GeV provide a
comparison at a higher collision energy.

\sect{Method}
\begin{figure*}[ht]
\centering
\includegraphics[width=1\textwidth]{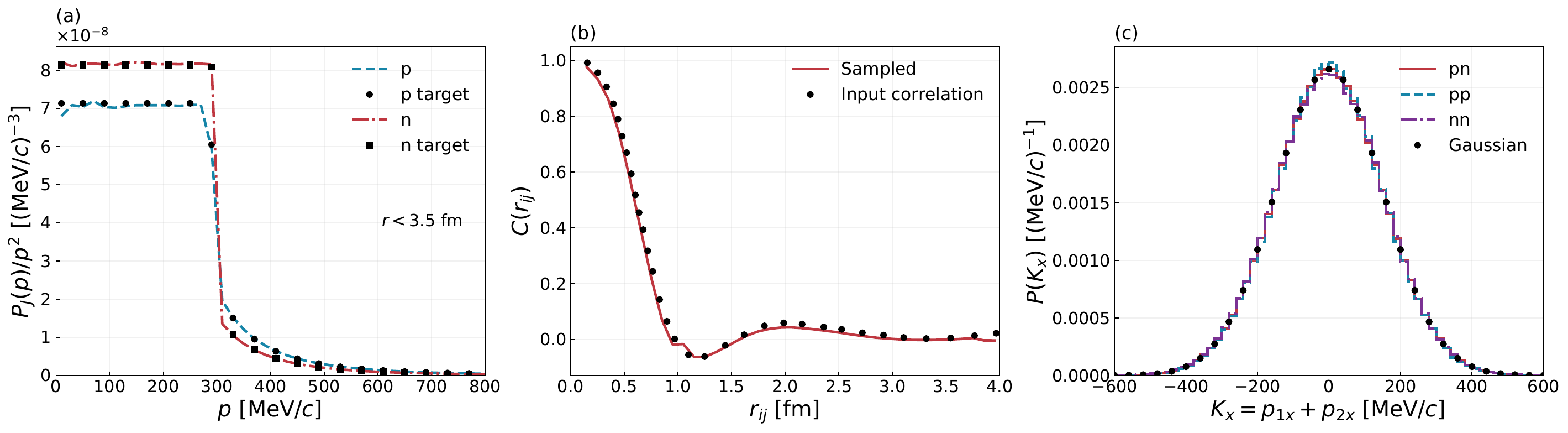}
\caption{(Color online)
Distributions of the initialized Au nuclei.
(a) Proton and neutron momentum distributions,
$P_J(p)/p^2$, within $r<3.5$ fm and their sampling targets.
(b) Final two-nucleon spatial correlation function
compared with the input.
(c) the $x$ component of the pair center-of-mass
momentum $K_x$ for designated
$pn$, $pp$, and $nn$ pairs, compared with a Gaussian
of width $150$ MeV/$c$.
}
\label{fig:src_distributions}
\end{figure*}

\subsection{Single-nucleon radial coordinate and momentum sampling}
\label{sec:momentum_sampling}

Each $^{197}$Au nucleus contains 118 neutrons and
79 protons. The one-body density follows a Woods--Saxon
profile,
\begin{equation}
\rho(r)=\frac{\rho_c}{1+\exp[(r-R)/a]},
\end{equation}
with $R=6.38$ fm and $a=0.535$ fm. The normalization $\rho_c$ is determined by
$4\pi\int_0^{15\,\mathrm{fm}}r^2\rho(r)\,dr=197$.

Unlike the species-dependent Skyrme--Hartree--Fock
densities used in Ref.~\cite{Liu:2026tde}, we assume
proportional neutron and proton density profiles:
\begin{equation}
\rho_n(r)=\frac{118}{197}\rho(r),
\qquad
\rho_p(r)=\frac{79}{197}\rho(r).
\end{equation}
The local isospin asymmetry is therefore spatially
uniform, $\delta=(\rho_n-\rho_p)/(\rho_n+\rho_p)=39/197$.

Radial coordinates are sampled with probability density
proportional to $r^2\rho(r)$, and an initial set of
118 neutron and 79 proton labels is assigned randomly.

 We adopt the single-nucleon momentum distribution and its default parameter set from Ref.~\cite{Liu:2026tde}.
The distribution includes a fermi-motion-like low-momentum
component and a high-momentum tail, as shown in fig.~\ref{fig:src_distributions}(a).

Following Ref.~\cite{Liu:2026tde}, each momentum magnitude is sampled using the nucleon's
radial coordinate and initial isospin label.
Nucleons sampled from the high-momentum tail are identified as high-momentum nucleons and marked for the subsequent pair construction.
The sampled radial coordinates, momentum magnitudes,
and branch indicators are retained during spatial
matching; isospin labels and momentum directions are
subsequently modified during the pair constructing stage as described below.

\subsection{Construction of correlated nucleon configurations}
\label{sec:correlated_configuration}

\subsubsection{Spatial correlations and pair matching}
\label{sec:spatial_matching}

For the sampled radial coordinates, we construct
spatially correlated configurations by assigning
nucleon position directions.
A candidate direction $\hat{\mathbf n}_i$ is sampled
isotropically, giving $\mathbf r_i=r_i\hat{\mathbf n}_i$.
The candidate is assigned a correlation weight
\begin{equation}
W_i=
\prod_{\substack{j<i\\r_{ij}<r_{\rm cut}}}
\left[1-C(r_{ij})\right],
\qquad
r_{ij}=|\mathbf r_i-\mathbf r_j|,
\end{equation}
where $j<i$ labels previously placed nucleons and
$r_{\rm cut}=3.8$ fm.
The spatial correlation function is adopted from
Refs.~\cite{Alvioli:2005cz,Alvioli:2009ab} and defined as
\begin{equation}
C(r)=1-\frac{\rho_{\rm corr}^{(2)}(r)}
               {\rho_{\rm uncorr}^{(2)}(r)},
\end{equation}
where $\rho_{\rm corr}^{(2)}$ and
$\rho_{\rm uncorr}^{(2)}$ are the radial two-body
densities of correlated and uncorrelated configurations.

The spatial-correlation sampling follows the method developed in our previous work~\cite{Huang2025NNCorrelations}. The correlation weight $W_i$ is used as the acceptance probability: a candidate position is accepted if a uniformly sampled random number $u\in[0,1)$ satisfies $u<W_i$. Otherwise, only its direction is resampled, while its radial coordinate remains fixed.

So far, the single-nucleon radial coordinates and momentum magnitudes have been sampled, and spatial correlations have been introduced. We next identify spatially close pairs among the high-momentum nucleons for the subsequent construction of pair-momentum correlations. 
With the sampled positions fixed, approximately 3.44\% of all nucleons can be included in high-momentum pairs satisfying
\begin{equation}
r_{ij}<r_{\rm pair}, \qquad r_{\rm pair}=1~\mathrm{fm}.
\end{equation}
These pairs are identified by maximizing the number of pairs without assigning any nucleon to more than one pair.
Inclusive electron-scattering analyses infer two-nucleon SRC probabilities of order 20\% in carbon and iron nuclei~\cite{Egiyan2006SRCProbabilities}, which we use as a qualitative reference for the paired-nucleon fraction in our initialization.

To increase the number of spatially close high-momentum pairs, we exchange position directions between unpaired high-momentum nucleons and low-momentum nucleons at similar radial distances. For an exchange between a high-momentum nucleon $j$ and a low-momentum nucleon $l$,
\begin{equation}
\mathbf r'_j=r_j\hat{\mathbf n}_l,
\qquad
\mathbf r'_l=r_l\hat{\mathbf n}_j,
\qquad
|r_j-r_l|<0.3~\mathrm{fm}.
\end{equation}

We first attempt a single exchange that brings an unpaired high-momentum nucleon within $r_{\rm pair}$ of another unpaired high-momentum nucleon. If no such exchange is available, we attempt two exchanges, each between an unpaired high-momentum nucleon and one member of a nearby low-momentum pair. The exchanges are retained only if the two high-momentum nucleons satisfy the pairing-distance criterion. 

This procedure preserves each nucleon's radial coordinate and momentum magnitude. Since the direction exchanges can modify the two-nucleon relative-distance distribution $\rho_{12}(\Delta r)$, this distribution is checked for the final configurations.
Approximately 31.47\% of all nucleons are sampled from the high-momentum tail. After the direction exchanges, nucleons in designated pairs account for 21.58\% of the total, while approximately 9.9\% remain unpaired. A high momentum alone does not uniquely identify membership in an SRC pair: finite-nucleus independent-particle calculations also contain momentum components above a characteristic Fermi momentum~\cite{Bisconti2007Momentum,Arrington:2011xs}.
The remaining high-momentum nucleons retain their sampled momentum magnitudes without imposed pair-momentum correlations.

\subsubsection{Pair-isospin assignment}

Motivated by the observed predominance of proton--neutron SRC pairs~\cite{CLAS:2018xvc}, we adopt a target pair-count ratio of $N_{pn}:N_{pp}:N_{nn}=90:5:5$. Keeping the designated pairs fixed, we adjust the proton/neutron labels to reproduce the target pair ratio across the ensemble.
The target ratio is approached collectively across designated pairs in the ensemble, rather than separately in each nucleus. Proton/neutron labels are reassigned using random selections of pairs and their members, with compensating relabeling among eligible unpaired nucleons in the same nucleus to preserve exactly 79 protons and 118 neutrons. These adjustments leave all nucleon positions, momentum magnitudes, and designated partners unchanged, as shown in fig.~\ref{fig:src_distributions}.

\subsubsection{Correlated momentum directions}

Motivated by measurements of the center-of-mass motion of SRC pairs~\cite{CLAS:2018qpc}, we adopt an isotropic Gaussian model for the pair center-of-mass momentum.
For each designated SRC pair, the center-of-mass and relative momenta are defined as
\begin{equation}
\begin{aligned}
\mathbf K=\mathbf p_1+\mathbf p_2,
\qquad
\mathbf k=\frac{\mathbf p_1-\mathbf p_2}{2},
\end{aligned}
\end{equation}
where \(\mathbf p_1\) and \(\mathbf p_2\) are the momenta of the two nucleons. Their previously sampled magnitudes, denoted by \(p_1=|\mathbf p_1|\) and \(p_2=|\mathbf p_2|\), are held fixed throughout the construction.

The three Cartesian components of the pair center-of-mass momentum are sampled independently from a Gaussian distribution with zero mean and a standard deviation of \(\sigma_K=150~\mathrm{MeV}/c\):
\(K_\mu\sim\mathcal N(0,\sigma_K^2),
\qquad \mu=x,y,z.\)

For a given center-of-mass momentum magnitude \(K\), compatibility with the fixed single-nucleon momentum magnitudes requires
\[
|p_1-p_2|\le K\le p_1+p_2.
\]
The presampled $\mathbf K$ vectors are assigned to designated pairs subject to the compatibility condition above. The presampled $\mathbf K$ vectors are first assigned
to compatible pairs within each nucleus. If necessary,
the assignment is extended across the sampled nuclei,
while the designated nucleon pairs remain unchanged. This assignment preserves the sampled single-nucleon momentum magnitudes and the complete set of $\mathbf K$ vectors, without resampling or truncation.

For each compatible assignment, we define \(\hat{\mathbf K}=\mathbf K/K\) and sample a unit vector \(\hat{\mathbf e}_{\perp}\) with a uniformly distributed azimuthal angle in the plane perpendicular to \(\hat{\mathbf K}\). The component of the first nucleon's momentum along \(\hat{\mathbf K}\) is
\begin{equation}
\begin{aligned}
q_{\parallel}
=
\frac{K^2+p_1^2-p_2^2}{2K}.
\end{aligned}
\end{equation}
The two momentum vectors are then constructed as
\begin{equation}
\begin{aligned}
\mathbf p_1
=
q_{\parallel}\hat{\mathbf K}
+
\sqrt{p_1^2-q{\parallel}^2}\,
\hat{\mathbf e}_{\perp},
\qquad
\mathbf p_2=\mathbf K-\mathbf p_1.
\end{aligned}
\end{equation}
By construction, these vectors satisfy
\begin{equation}
\begin{aligned}
|\mathbf p_1|=p_1,
\qquad
|\mathbf p_2|=p_2,
\qquad
\mathbf p_1+\mathbf p_2=\mathbf K.
\end{aligned}
\end{equation}
Unpaired nucleons retain independently sampled isotropic momentum directions.

\begin{figure*}[t]
\centering
\includegraphics[width=\textwidth]{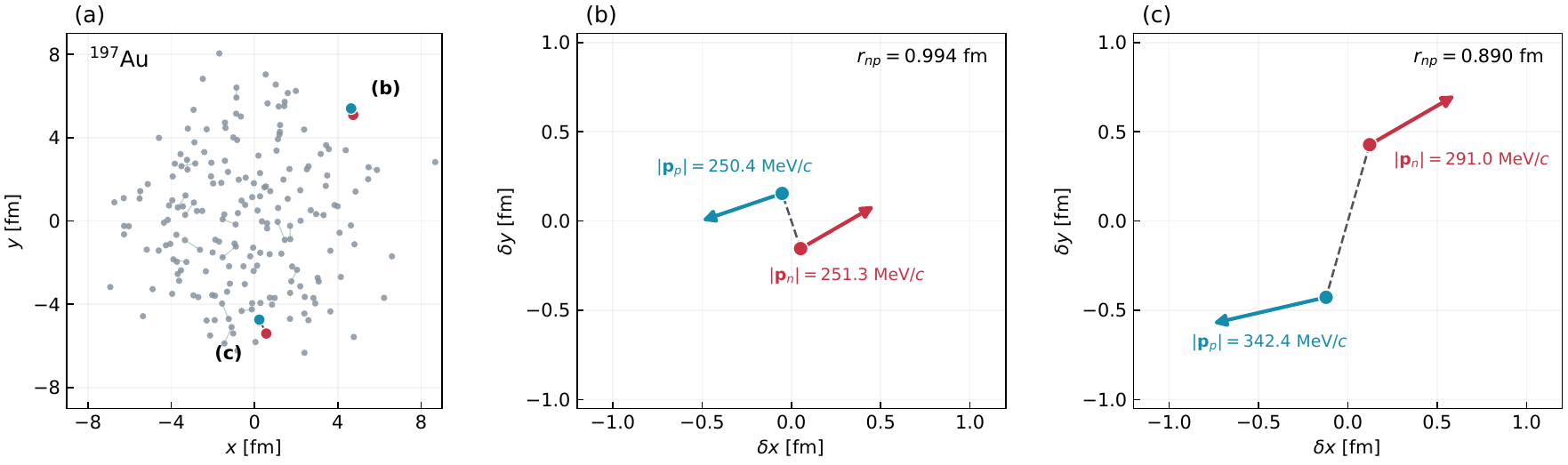}
\caption{(Color online)
Two selected high-momentum $pn$ pairs in a sampled
$^{197}$Au nucleus.
(a) Nuclear spatial projection with the pairs highlighted
and identified by their corresponding panel labels.
(b,c) Pair configurations relative to their midpoints,
projected onto their respective momentum planes.
Red and blue arrows represent neutron and proton momenta.
}
\label{fig:src_example}
\end{figure*}

Figure~\ref{fig:src_example} illustrates the spatial
configurations and individual nucleon momenta of two
selected high-momentum $pn$ pairs from the same
initialized nucleus.
Figure~\ref{fig:src_distributions} summarizes the
distributions after the full initialization procedure.
Panel (a) compares the proton and neutron momentum
distributions within $r<3.5$ fm with their sampling
targets; panel (b) shows the final two-nucleon spatial
correlation function; and panel (c) compares the $x$ component of the pair center-of-mass
momentum, $K_x$, with the input Gaussian distribution.

\subsection{Transport evolution with SMASH}
\label{sec:smash}

We employ the hadronic transport model SMASH 3.4.1
to simulate the dynamical evolution of Au+Au collisions.
SMASH describes nonequilibrium hadronic systems through
hadron propagation, scattering, resonance formation,
and decay~\cite{SMASH:2016zqf,Steinberg:2018jvv}.

The particle evolution includes mean-field propagation and discrete collision processes. Schematically, the single-particle distribution satisfies
\[
\frac{\partial f_i}{\partial t}
+\nabla_{\mathbf p}H_i\cdot\nabla_{\mathbf r}f_i
-\nabla_{\mathbf r}H_i\cdot\nabla_{\mathbf p}f_i
=I_i^{\mathrm{coll}},
\]
where \(H_i\) denotes the single-particle Hamiltonian including the mean-field contribution, and \(I_i^{\mathrm{coll}}\) describes scattering, resonance formation, and decay processes.

The isoscalar single-particle potential is parametrized
as \cite{Welke:1988zz,Tarasovicova:2024isp}
\[
U(\rho,p)=Ax+Bx^\tau+
\frac{2C}{\rho_0}\int d^3q\,
\frac{F(\rho,q)}
{1+|\mathbf p-\mathbf q|^2/(\hbar c\Lambda)^2},
\]
where \(x=\rho/\rho_0\),\(\rho_0=0.168\ {\rm fm}^{-3}\) and
\(\int F(\rho,q)d^3q=\rho\). \(F(\rho,q)\) in the mean-field integral is approximated by a cold Fermi distribution.

In this study, SRC effects are incorporated primarily through the input nucleon phase-space configurations and their subsequent transport evolution. 
For the SRC ensemble, the sampled nucleon positions,
three-momenta, and isospin labels are supplied directly
as external initial conditions. A common random global rotation is
applied to both positions and momenta, nucleon energies are
determined by the on-shell relation, and internal Fermi-momentum
sampling is bypassed. 
We employ
frozen-Fermi propagation and switch off the mean field during
the negative-time stage. 
At $t=0$ fm/c, the freezing is lifted and the mean field is
activated using the current particle distribution.
The system is then evolved to $t=50$ fm/$c$.

We simulate Au+Au collisions at $\sqrt{s_{NN}}=2.4$
and 3.2 GeV. At 2.4 GeV, three calculations are performed:
WS initialization with the default mean field, SRC
initialization with the same default mean field, and
SRC initialization with a refitted mean field.
Each calculation comprises 200,000 generated events
with impact parameters $0\leq b\leq15$ fm.
After excluding events without interactions, the
corresponding samples contain 187,070, 198,205,
and 197,216 events, respectively.
Centrality classes are determined separately for each
sample from its final-state charged-particle multiplicity
distribution.
At $\sqrt{s_{NN}}=3.2$~GeV, the SRC and WS calculations each comprise 200,000 generated events and use the same momentum-independent mean field, with $A=-209.2$~MeV, $B=156.4$~MeV, $\tau=1.35$, and $S_{\rm pot}=18$~MeV, corresponding to an incompressibility of $K\simeq240$~MeV.

The WS reference uses the SMASH default with Woods-Saxon positions and Fermi momenta. 
The mean field acts from the initialization stage onward.
Pauli blocking is implemented with a spatial averaging radius of \(1.86~\mathrm{fm}\), a momentum averaging radius of \(0.08~\mathrm{GeV}/c\), and a Gaussian cutoff parameter of \(2.2~\mathrm{fm}\).
Proton anisotropic flow is evaluated relative to the
reaction plane as
\(v_n=\langle\cos[n(\phi-\Psi_{\mathrm{RP}})]\rangle\),
where \(\phi\) is the proton azimuthal angle and
\(\Psi_{\mathrm{RP}}=0\) in the simulation coordinates.
The coefficients \(v_1\) and \(v_2\) denote directed
and elliptic flow, respectively.

\subsection{Mean-field recalibration with SRC initialization}

For the SRC initialization, we vary only the mean-field parameters while keeping the initialization prescription and all other transport settings fixed. 

The mean field parameters $(K,C,\Lambda,S_{\rm pot})$
are adjusted using charged-pion multiplicities and
differential proton elliptic flow~\cite{Danielewicz:1999zn}.
For each choice of $(K,C,\Lambda)$, the parameters
$(A,B,\tau)$ are determined by the saturation
binding energy, vanishing pressure at saturation,
and specified incompressibility.
Candidate selection uses Gaussian-process emulators
and additional nuclear-matter screening criteria,
as described in Appendix~\ref{app:mf_calibration}.

\sect{Results}
\subsection{Charged-pion production and proton flow at \(\sqrt{s_{NN}}=2.4\) GeV}

We first compare the WS and SRC results with the same
default mean field and then examine the changes induced
by refitting the mean field for SRC initialization.

\begin{figure*}[htp]
\centering
\includegraphics[width=1\textwidth]{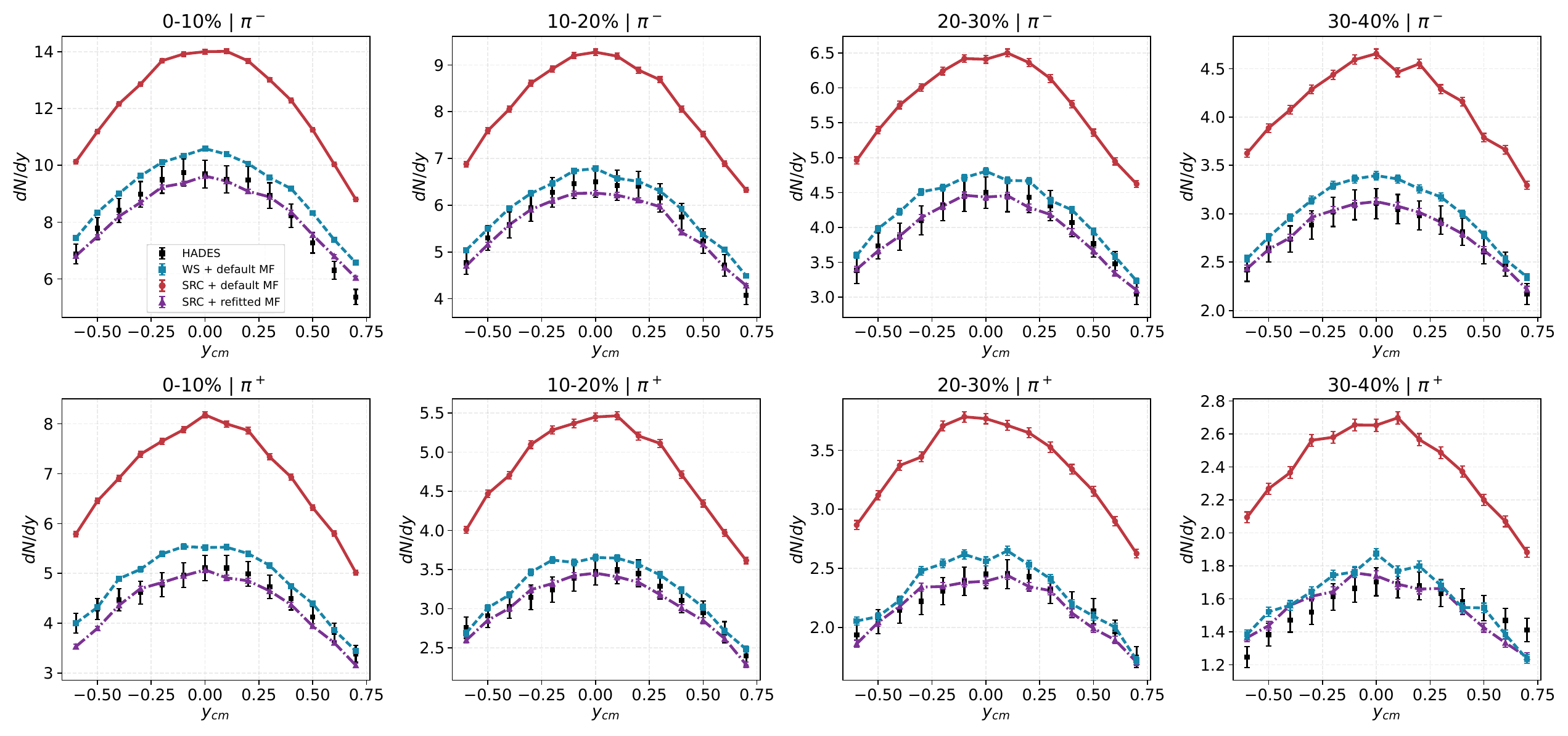}
\caption{(Color online)
Rapidity densities of $\pi^-$ (upper panels) and
$\pi^+$ (lower panels) in Au+Au collisions at
$\sqrt{s_{NN}}=2.4$ GeV for the indicated centrality
classes.
Black symbols denote HADES data
\cite{HADES:2020ver}.
}
\label{fig:dndy2p4}
\end{figure*}

Figure~\ref{fig:dndy2p4} shows the rapidity distributions of
$\pi^-$ and $\pi^+$ in the 0--10\%, 10--20\%, 20--30\%,
and 30--40\% centrality classes. All three calculations yield distributions that peak near midrapidity and decrease toward forward and backward rapidities. The yields of both pion species decrease from central to peripheral collisions. With the default mean field, SRC initialization produces higher yields than the WS reference. For example, in the 0–10\% centrality class and within \(|y_{\rm cm}|<0.05\), the \(\pi^-\) and \(\pi^+\) rapidity densities increase by approximately 34\% and 47\%, respectively, relative to the WS calculation. The enhancement therefore affects both charge states, with a larger relative response for \(\pi^+\). Mean-field refitting reduces both pion yields in the SRC calculation, bringing their overall magnitudes closer to the WS reference and the experimental data while preserving the general rapidity and centrality dependence.
Reduced transverse-mass spectra of $\pi^-$ and $\pi^+$ in
0--10\% central collisions, at $|y_{\rm cm}|<0.05$ and
$0.65<y_{\rm cm}<0.75$, are presented in
Appendix~\ref{app:pion_mt}.

\begin{figure*}[htp]
\centering
\includegraphics[width=1\textwidth]{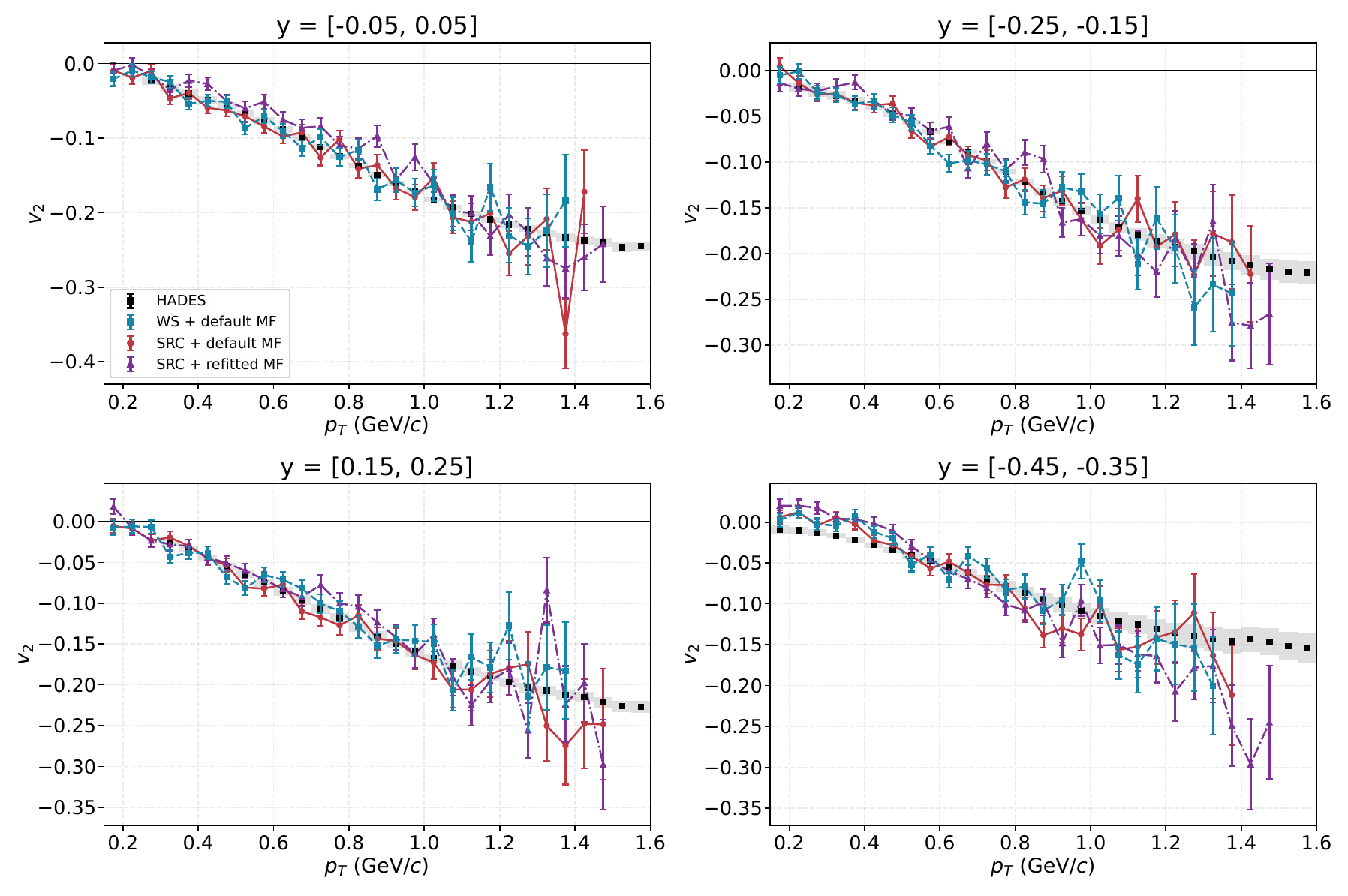}
\caption {\label{fig:v2pt2p4}(Color online) Proton elliptic flow \(v_2\) as a function of transverse momentum in 20–30\% central Au+Au collisions at \(\sqrt{s_{NN}}=2.4\) GeV, for the four center-of-mass rapidity intervals indicated in the panels. SRC and WS results are compared with HADES measurements \cite{HADES:2022osk}.
}
\end{figure*}

In addition to charged-pion multiplicities, proton elliptic
flow $v_2(p_T)$ is used to refit the mean-field parameters
for the SRC initialization.
Figure~\ref{fig:v2pt2p4} shows $v_2(p_T)$ in 20--30\%
central Au+Au collisions at $\sqrt{s_{NN}}=2.4$ GeV
for four rapidity intervals.
All three calculations exhibit a similar transverse-momentum
dependence, with $v_2$ generally becoming more negative as
$p_T$ increases.
Compared with the substantial change in pion yields,
the mean-field refit produces a smaller change in the
displayed proton elliptic flow.
Additional proton-flow results are presented in
Appendix~\ref{app:additional_flow}.

The refit changes both the density and momentum dependence
of the mean field. Compared with the default parameter set
listed in Table~\ref{table:coefficients}, the SRC-refitted parametrization has a
lower incompressibility, a larger momentum scale $\Lambda$,
and a larger magnitude of the negative momentum-dependent
coupling $C_{\rm MD}$.

\begin{figure*}[htp]
\centering
\includegraphics[width=0.6\textwidth]{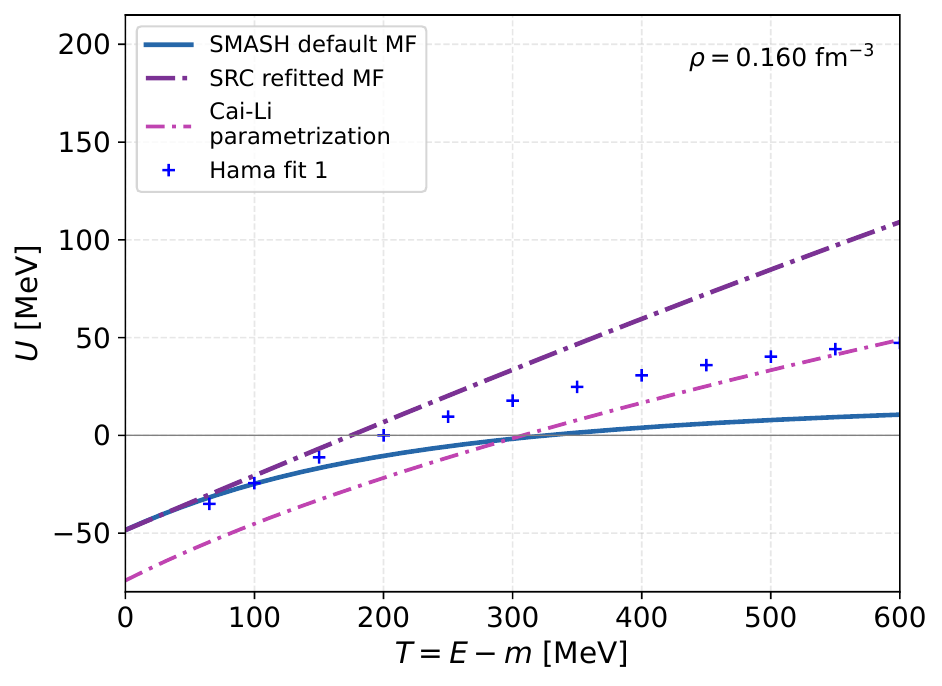}
\caption{(Color online)
Single-particle potentials at $\rho=0.160$ fm$^{-3}$
for the default, refitted, and Cai--Li parametrizations.
Blue plus signs denote the Schrödinger-equivalent
optical reference reconstructed from
Ref.~\cite{Hama:1990vr}.
}
\label{fig:optical}
\end{figure*}

Figure~\ref{fig:optical} compares the resulting
single-particle potentials with the Cai--Li
parametrization incorporating an experimentally
constrained high-momentum tail~\cite{Cai:2022grw}
and the optical reference reconstructed from
Ref.~\cite{Hama:1990vr}.
Over the displayed kinetic-energy range, the refitted
potential rises more steeply with energy and becomes
substantially more repulsive toward the upper end of
the interval.

\begin{table}[t]
\centering
\caption{Mean-field parameters for Au+Au collisions at
$\sqrt{s_{NN}}=2.4$ GeV. The SRC-refitted parametrization
retains the cold Fermi-sphere approximation in the
momentum-dependent mean-field integral.}
\label{table:coefficients}
\begin{tabular}{lrr}
\hline\hline
Parameter & SMASH default & SRC refitted \\
\hline
$K$ (MeV) & 348.20 & 214.78 \\
$A$ (MeV)              & $-15.436746$ & $186.223697$  \\
$B$ (MeV)              & $42.579160$  & $403.779145$  \\
$\tau$                 & $2.174899$   & $1.091261$    \\
$C_{\mathrm{MD}}$ (MeV)& $-63.105235$ & $-334.262842$ \\
$\Lambda$ (fm$^{-1}$)  & $2.119523$   & $8.609508$    \\
$S_{\mathrm{pot}}$ (MeV)& $15.8$      & $13.650655$   \\
\hline\hline
\end{tabular}
\end{table}

\begin{figure*}[htp]
\centering
\includegraphics[width=0.6\textwidth]{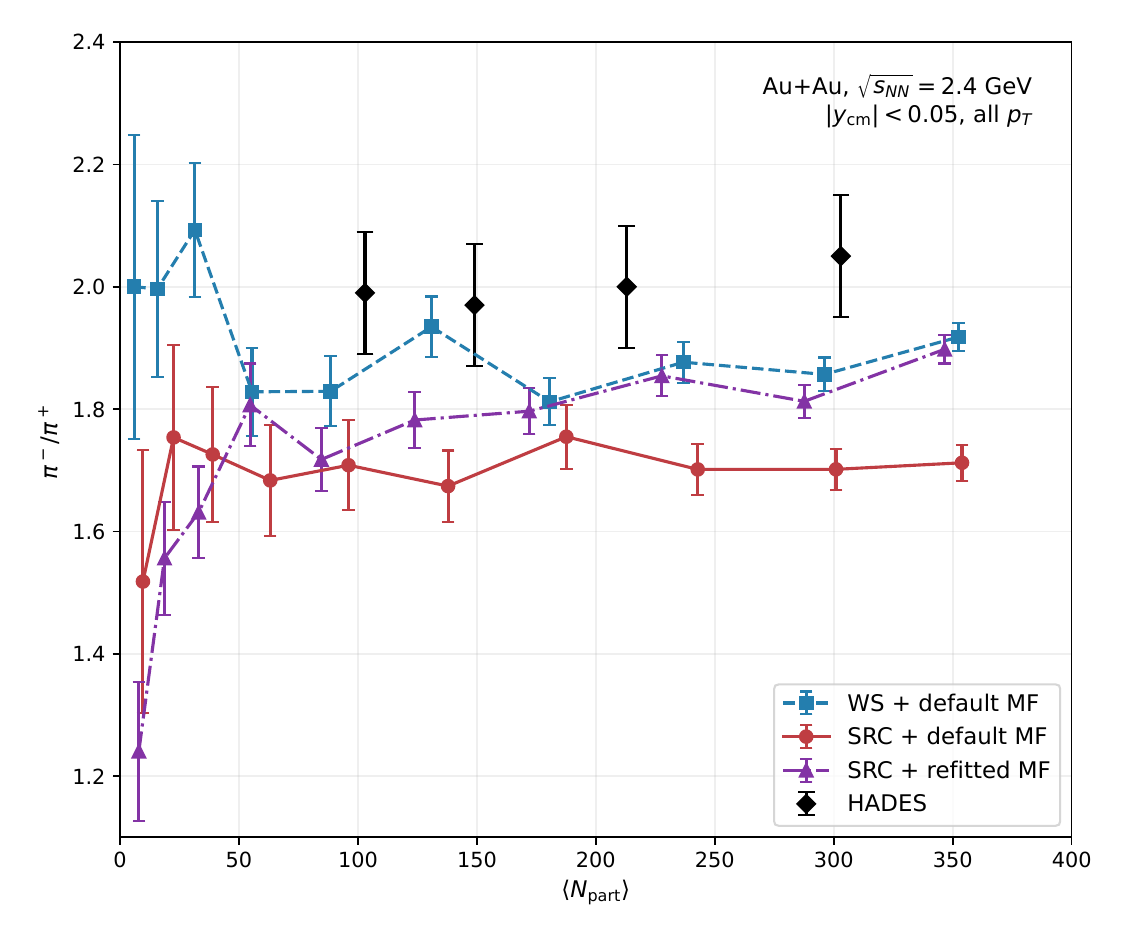}
\caption {\label{fig:ratio2p4}(Color online) Charged-pion yield ratio \(R_\pi=\sum N_{\pi^-}/\sum N_{\pi^+}\) within \(|y_{\mathrm{cm}}|<0.05\) in Au+Au collisions at \(\sqrt{s_{NN}}=2.4\) GeV. SRC and WS results are plotted against the model participant-number proxy \(\langle N_{\mathrm{part}}^{\mathrm{proxy}}\rangle=\langle394-N_{n_{\mathrm{coll}}=0}^{p,n}\rangle\), where \(N_{n_{\mathrm{coll}}=0}^{p,n}\) counts final-state protons and neutrons without recorded collisions. Black diamonds denote HADES experimental data \cite{HADES:2022mwn}. 
}
\end{figure*}

Figure~\ref{fig:ratio2p4} shows the midrapidity $\pi^-/\pi^+$ ratio integrated over the full transverse-momentum range. With the default mean field, the SRC calculation gives a lower charged-pion ratio than the WS reference, consistent with the larger relative enhancement of $\pi^+$ production. At larger participant numbers, the refitted SRC calculation gives a higher ratio than the SRC calculation with the default mean field and lies close to the WS result. At the smallest participant numbers shown, both SRC calculations give lower ratios than the WS calculation.

\subsection{Disentangling the contributions of SRC initialization ingredients}

The SRC and WS initializations differ in their spatial
configurations, single-nucleon momentum distributions,
pair-isospin composition, and pair-momentum structure.
To identify the origin of the pion-yield enhancement,
Table~\ref{table:initial_configuration} compares five
initial configurations for Au+Au collisions at
$\sqrt{s_{NN}}=2.4$ GeV and $b=0$ fm, with 10,000 events
per configuration.
A and B differ in their pair-isospin assignments,
whereas B and C differ in their initial momenta.
D provides the Woods--Saxon reference with Fermi momenta.
Configuration E uses the same single-nucleon momentum distribution as B, but replaces its correlated spatial configuration with Woods--Saxon positions and removes the original position--momentum, isospin--momentum, and pair associations.
The B--C comparison demonstrates this momentum dependence, whereas changing the pair-isospin composition between A and B alters both yields by less than 2\%. The enhancement persists in E even after the original spatial and pair associations are removed. These controlled comparisons identify the high-momentum tail of the initial single-nucleon distribution as the dominant source of the enhancement in both $\pi^+$ and $\pi^-$ yields.

\begin{table*}[t]
\centering
\caption{
Mean charged-pion multiplicities and their ratio for five
initialization configurations in Au+Au collisions at
$\sqrt{s_{NN}}=2.4$ GeV and $b=0$.
Each configuration contains 10,000 events.
}
\label{table:initial_configuration}

\small
\setlength{\tabcolsep}{5pt}
\renewcommand{\arraystretch}{1.15}

\begin{tabular}{clllrrr}
\toprule
Config. &
Positions &
Momenta &
\shortstack{Isospin\\assignment} &
$\langle N_{\pi^+}\rangle$ &
$\langle N_{\pi^-}\rangle$ &
$R_\pi$ \\
\midrule
A & SRC & SRC
  & $pn:pp:nn=90:5:5$
  & 17.0456 & 28.8403 & 1.6919 \\

B & SRC & SRC
  & Random
  & 17.3380 & 28.9645 & 1.6706 \\

C & SRC & Fermi momenta
  & Random
  & 11.8996 & 21.6873 & 1.8225 \\

D & Woods--Saxon & Fermi momenta
  & Random
  & 11.7868 & 21.6492 & 1.8367 \\

E & Woods--Saxon & High-momentum tail
  & Random
  & 16.7667 & 28.2157 & 1.6828 \\
\bottomrule
\end{tabular}
\end{table*}

To trace the influence of the initial momentum configuration on the charged-pion ratio, we analyzed the complete reaction histories of two samples containing 10000 events each. The samples share the same SRC spatial configurations, random proton/neutron labels, and simulation settings, differing only in their initial momenta: SRC momenta in one sample and Fermi momenta sampled by SMASH in the other. Both samples use identical propagation settings and are evolved to $t=50$ fm/$c$.
The corresponding time evolution of particle
multiplicities and ratios is presented in
Appendix~\ref{app:optical}.

\begin{table*}[t]
\centering
\caption{Mean cumulative numbers of accepted $NN\to N\Delta(1232)$ and $NN\to\Delta(1232)\Delta(1232)$ reactions per event in the matched Fermi-momentum and SRC-momentum config. C and B, respectively, each containing 10,000 events at $b=0$ with the default mean field. The increase is the SRC mean minus the Fermi mean, and the relative increase is evaluated with respect to the Fermi sample.}
\label{table:reaction_table}
\begingroup
\setlength{\tabcolsep}{12pt}
\renewcommand{\arraystretch}{1.2}
\begin{tabular}{lcccc}
\hline
Reaction channel
& \shortstack{Fermi momenta\\(C)}
& \shortstack{SRC momenta\\(B)}
& \shortstack{Mean increase\\(B$-$C)}
& \shortstack{Relative increase\\(\%)} \\
\hline
$pp\to N\Delta$     & 23.0786 & 26.4510 & 3.3724 & 14.61 \\
$pn\to N\Delta$     & 36.6569 & 40.6629 & 4.0060 & 10.93 \\
$nn\to N\Delta$     & 43.4829 & 47.6814 & 4.1985 & 9.66 \\
$pp\to\Delta\Delta$ & 0.2519  & 0.6647  & 0.4128 & 163.87 \\
$pn\to\Delta\Delta$ & 1.7435  & 3.9653  & 2.2218 & 127.43 \\
$nn\to\Delta\Delta$ & 0.5876  & 1.3085  & 0.7209 & 122.69 \\
\hline
\end{tabular}
\endgroup
\end{table*}

Table~\ref{table:reaction_table} shows that SRC momenta enhance resonance production in all six channels, with substantially larger relative increases for $NN\to\Delta\Delta$ than for $NN\to N\Delta$. Summed over the incoming charge channels, the mean cumulative reaction counts per event increase by 11.2\% and 129.9\% for $NN\to N\Delta$ and $NN\to\Delta\Delta$, respectively. The $pp$ channels exhibit the largest relative increases in both cases. 
The time evolution in Appendix~\ref{app:optical}
also shows a larger proton-to-neutron ratio in the
SRC-momentum sample.
Charge-conserving reaction chains such as
$nn\to p\Delta^-\to pn\pi^-$ convert one neutron into a proton while producing a negative pion, thereby increasing the proton-to-neutron ratio.
This change in nucleon composition may contribute to the enhanced $pp$ reaction counts. The stronger relative enhancement of the $pp$ channels is consistent with the larger relative increase in the final $\pi^+$ yield (45.7\%) than in the $\pi^-$ yield (33.5\%).

\subsection{Mean-field compensation of the SRC-induced pion enhancement}



Variations of $K$ and $S_{\rm pot}$ at fixed momentum-dependent parameters do not remove the pion-yield excess in the tested calculations (Appendix~\ref{app:KS_scan}). We therefore compare three calculations with the same SRC initial states but different strengths of high-momentum repulsion to examine how the momentum-dependent mean field affects pion production.

\begin{figure*}[htp]
\centering
\includegraphics[width=1\textwidth]{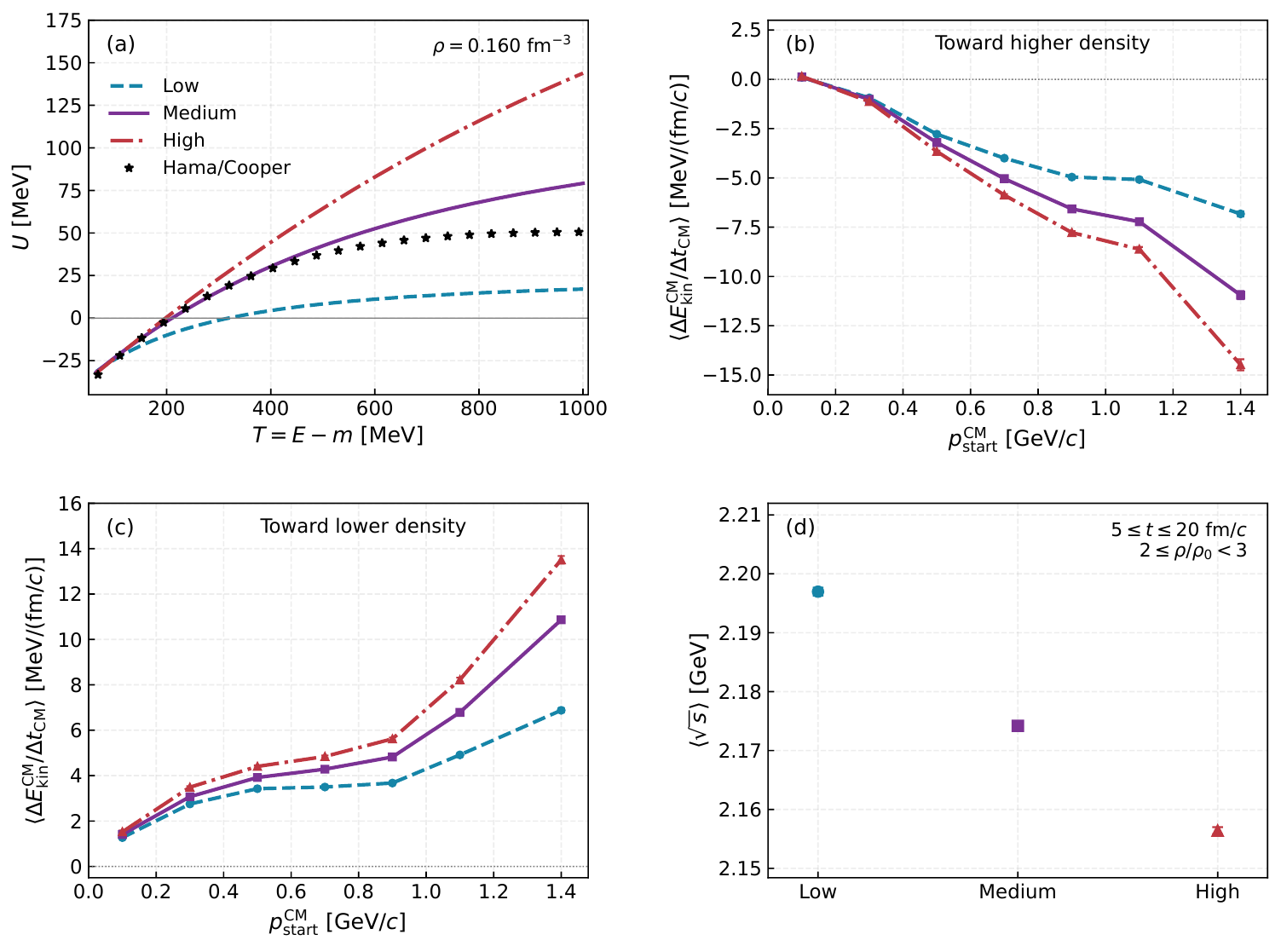}
\caption{
Effects of varying the high-momentum mean field:
(a) single-particle potentials and the optical reference;
(b,c) mean kinetic-energy change rates
$\langle q_{\mathrm{kin}}\rangle$ during collision-free
propagation toward higher and lower density;
(d) mean two-particle center-of-mass energy
$\langle\sqrt{s}\rangle$ of accepted $NN$ collisions.
Panels (b,c) select $\bar{\rho}\geq\rho_0$, while
panel (d) selects $2\leq\rho/\rho_0<3$;
both analyses use $5\leq t\leq20$ fm/$c$.
Directions in (b,c) are defined in the local baryon
rest frame, while the plotted momenta and energy-change
rates are evaluated in the nucleus--nucleus
center-of-mass frame.
}
\label{fig:high_p}
\end{figure*}

Figure \ref{fig:high_p} examines the response of pion production to the
high-momentum mean field using three calculations with
the same SRC initial states and matched event settings.
The incompressibility and symmetry-potential strength
are fixed at $K=240$ MeV and $S_{\mathrm{pot}}=15.8$ MeV. The single-particle potentials remain
relatively close at low energies but increasingly
separate at higher energies [Fig. \ref{fig:high_p}(a)].
With 1000 events per setting, the mean final-state $\pi^+$ multiplicities are 8.072, 6.498, and 5.505 for the Low, Medium, and High settings, respectively, while the corresponding $\pi^-$ multiplicities are 14.088, 11.792, and 10.200. Both yields decrease monotonically with increasing high-momentum repulsion, with reductions of 31.8\% for $\pi^+$ and 27.6\% for $\pi^-$ from Low to High.

To examine how the mean field modifies nucleon kinetic energies between collisions, we evaluate their rates of change over collision-free propagation intervals $[t_0,t_1]$ during the compression stage, $5\leq t_0<t_1\leq20$~fm/$c$. We select intervals with a mean endpoint density $\bar{\rho}=[\rho(t_0)+\rho(t_1)]/2\geq\rho_0$, where the density is reconstructed using the native SMASH density prescription and lattice geometry. For each interval, we calculate
\begin{equation}
\begin{aligned}
q_{\mathrm{kin}}
&=
\frac{E_{\mathrm{kin}}^{\mathrm{CM}}(t_1)
      -E_{\mathrm{kin}}^{\mathrm{CM}}(t_0)}
     {t_1-t_0},
\\
E_{\mathrm{kin}}^{\mathrm{CM}}(t)
&=
\sqrt{m_N^2+|\mathbf p^{\mathrm{CM}}(t)|^2}-m_N ,
\end{aligned}
\end{equation}
where $c=1$ and CM denotes the nucleus--nucleus
center-of-mass frame.

Figures~\ref{fig:high_p}(b) and \ref{fig:high_p}(c)
show $\langle q_{\mathrm{kin}}\rangle$, the mean
kinetic-energy change rate over selected propagation
intervals in each $p_{\mathrm{start}}^{\mathrm{CM}}$
bin.
Here, $p_{\mathrm{start}}^{\mathrm{CM}}
=|\mathbf p^{\mathrm{CM}}(t_0)|$ is the momentum
at the beginning of each interval.
The two panels distinguish motion toward higher and
lower density, respectively, with directions classified
relative to the density gradient in the local baryon
rest frame.
We distinguish motion toward higher and lower density by the angle $\theta$ between the nucleon's displacement over the propagation interval and the mean density gradient, both evaluated in the local baryon rest frame. Since the gradient points toward increasing density, intervals with $\cos\theta>0.3$ are classified as motion toward higher density, while those with $\cos\theta<-0.3$ are classified as motion toward lower density.
For nucleons moving toward higher density, the mean
kinetic-energy change rate becomes more negative from
Low to High, particularly at larger momenta.
For motion toward lower density, the corresponding rate
is positive and increases with the high-momentum repulsion.
Thus, stronger repulsion can enhance kinetic-energy loss
during motion toward denser matter while enhancing the
gain during outward motion.

Figure~\ref{fig:high_p}(d) shows $\langle\sqrt{s}\rangle$, averaged over all accepted $NN$ collisions satisfying these selections, including elastic and inelastic channels.
Here, $\sqrt{s}=\sqrt{(p_1^\mu+p_2^\mu)^2}$ is calculated
from the incoming nucleon four-momenta immediately before
each accepted collision.
The mean decreases from approximately $2.1970$ GeV
for the Low setting to $2.1564$ GeV for the High
setting, a reduction of about $40.6$ MeV.
The lower collision energies reduce the phase space
available for $NN\to N\Delta$, since the accessible
resonance mass is bounded by
$m_\Delta\leq\sqrt{s}-m_N$.
Near $m_N+m_\Delta^{\mathrm{pole}}\simeq2.17$ GeV,
this reduction restricts access to the spectral strength
around the $\Delta$ pole and can suppress resonance
production.

A more repulsive nucleon mean field at high momenta is thus accompanied by a lower mean kinetic invariant energy of accepted \(NN\) reactions in the selected compression-stage window and fewer \(NN\rightarrow N\Delta\) production reactions. These changes provide a dynamical explanation for the reduced pion yields in Figure \ref{fig:high_p}, showing how mean-field adjustments can compensate for the enhancement generated by the SRC momentum initialization.
The refitted parameters are used here to characterize this compensation within the adopted transport setup, rather than as an independent determination of the empirical optical potential.

\subsection{Charged-kaon production}

\begin{figure*}[htp]
\centering
\includegraphics[width=1\textwidth]{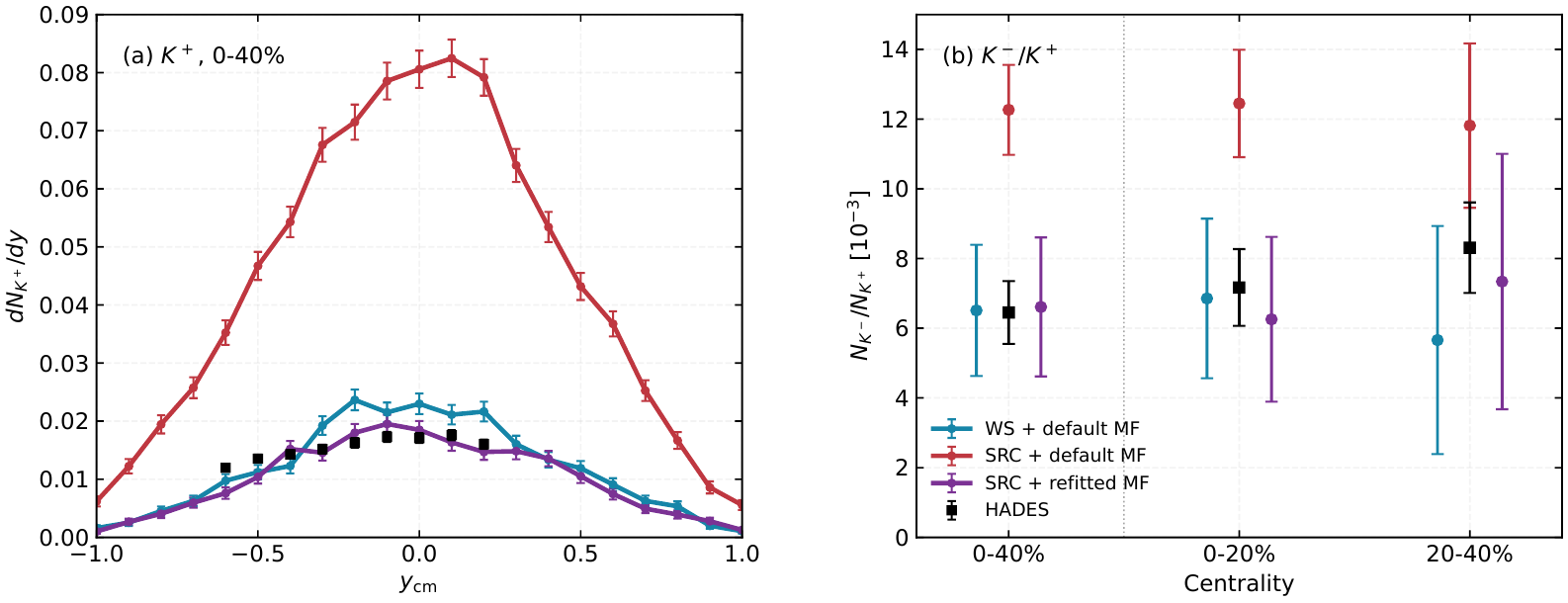}
\caption{
(Color online) Charged-kaon production in Au+Au
collisions at $\sqrt{s_{NN}}=2.4$~GeV.
(a) $K^+$ rapidity distributions in the 0--40\%
centrality class.
(b) $K^-/K^+$ ratios in the indicated centrality
intervals.
Results for WS initialization with the default mean
field and SRC initialization with the default and
refitted mean fields are compared with HADES data.
The experimental points in panel (a) are digitized
from the filled symbols in the right panel of
Fig.~5.21 of Ref.~\cite{Schuldes2016Kaons};
those in panel (b) are taken from Table~I of
Ref.~\cite{HADES:2017jgz}.
}
\label{fig:kaon_rapidity_ratio}
\end{figure*}

Figure~\ref{fig:kaon_rapidity_ratio}(a) shows the $K^+$
rapidity distributions in 0--40\% central Au+Au collisions
at $\sqrt{s_{NN}}=2.4$~GeV. With the default mean field,
the SRC calculation gives substantially higher yields
than the WS reference throughout the displayed rapidity
range and overestimates the HADES data. The SRC
calculation with the refitted mean field gives a much
lower yield, bringing the rapidity distribution closer
to the measurements. This qualitative response is similar to that of
charged-pion production. 
Using the same $b=0$ samples as in Table~\ref{table:reaction_table}, together with the corresponding calculations using the refitted mean field, we find reduced production of $\Delta(1232)$ resonances and of heavier $N^*$ and $\Delta^*$ resonances feeding strange decay channels. These changes support reduced resonance production as a common contribution to the suppression of pion and $K^+$ yields.

At this collision energy, $K^-$ production lies further
below its free nucleon--nucleon threshold than $K^+$
production. 
In the 0--40\% centrality class, the mean $K^-$
multiplicities are approximately $1.6\times10^{-4}$,
$1.1\times10^{-3}$, and $1.4\times10^{-4}$ per event
for the WS-default, SRC-default, and SRC-refitted
calculations, respectively. Their ordering follows
the same qualitative pattern as the $K^+$ yields.

Figure~\ref{fig:kaon_rapidity_ratio}(b) compares the
$K^-/K^+$ ratios in three centrality intervals.
The SRC-default calculation gives ratios above the
WS reference and the experimental values, in contrast
to the lower $\pi^-/\pi^+$ ratio obtained with SRC
initialization. The WS-default and SRC-refitted kaon
ratios are compatible with the data within their
uncertainties. The WS-default and SRC-refitted ratios remain close within the relatively large statistical uncertainties associated with the small $K^-$ yields.

\subsection{Au+Au collisions at \(\sqrt{s_{NN}}=3.2~\mathrm{GeV}\)}

We next examine the initialization dependence at
$\sqrt{s_{NN}}=3.2$ GeV using the momentum-independent
mean field specified in Sec.~\ref{sec:smash}.

\begin{figure*}[htp]
\centering
\includegraphics[width=1\textwidth]{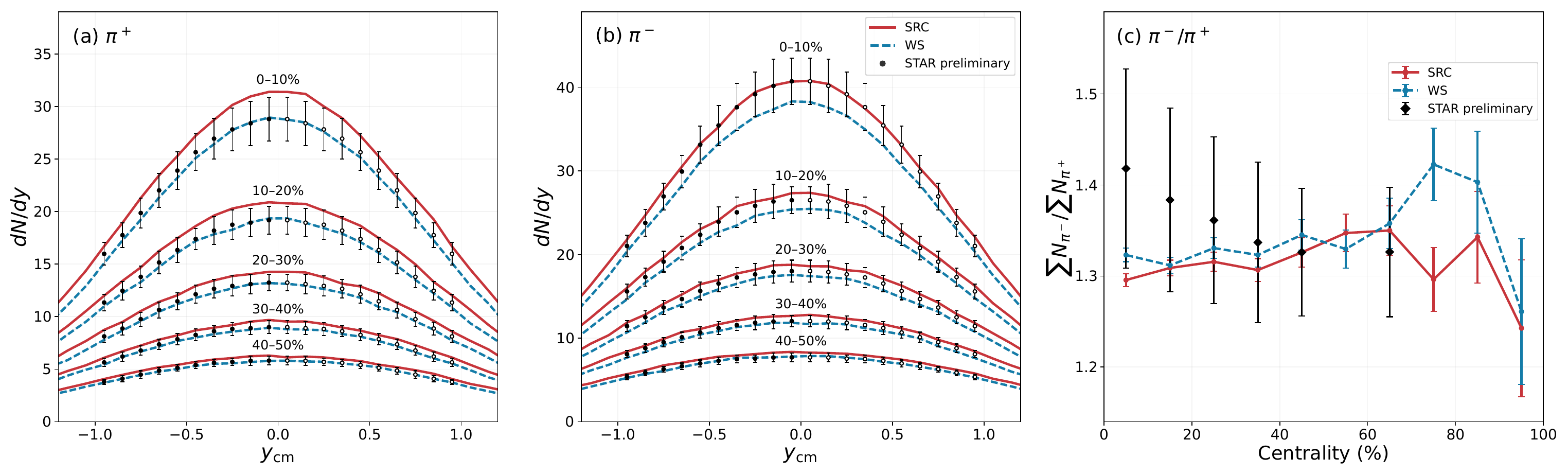}
\caption{(Color online)
Charged-pion production in Au+Au collisions at
$\sqrt{s_{NN}}=3.2$ GeV.
(a,b) $\pi^+$ and $\pi^-$ rapidity densities for
the indicated centrality classes.
(c) Charged-pion multiplicity ratio within
$-0.1<y_{\mathrm{cm}}<0$, integrated over all $p_T$.
Red and blue curves denote the SRC and WS calculations,
respectively, both using a mean field without explicit
momentum dependence. 
Black symbols denote STAR preliminary data
\cite{Chen_STAR_CPOD2026}.
}
\label{fig:3p2gev_pions}
\end{figure*}

Panels (a) and (b) of Fig.~\ref{fig:3p2gev_pions}
show the $\pi^+$ and $\pi^-$ rapidity densities,
respectively, for five centrality classes spanning
0--50\%. Throughout the displayed rapidity range, the SRC sample yields more pions of both charges than the WS sample, with a larger relative enhancement for \(\pi^+\). For the 0--10\% centrality class, the SRC $\pi^+$
yield near midrapidity lies above the STAR preliminary
central values, whereas the WS yield is close to them.
For $\pi^-$ in the same centrality class, the SRC
result is close to the experimental central values,
while the WS result lies below them.

The right panel of Figure \ref{fig:3p2gev_pions}  presents the centrality dependence of the charged-pion ratio, integrated over all transverse momenta within the midrapidity window \(-0.1<y_{\mathrm{cm}}<0\). Over 0–50\% centrality, both ratios lie approximately between 1.30 and 1.35, with the SRC central values below those of WS; the difference is small in the 10–20\% interval. Their ordering reverses in the 50–60\% interval, after which the SRC central values again fall below those of WS. Statistical uncertainties increase toward peripheral collisions, however, preventing a firm inference of the peripheral trend from the central values alone. The STAR preliminary central values decrease from approximately 1.42 in central collisions to approximately 1.33 in more peripheral collisions. Over 0–40\% centrality, both model central values lie below the corresponding experimental values.

These results indicate that SRC initialization continues to enhance charged-pion production in Au+Au simulations at \(\sqrt{s_{NN}}=3.2~\mathrm{GeV}\), while reducing \(\pi^-/\pi^+\) in central collisions through the larger relative enhancement of the \(\pi^+\) yield. The enhancement common to both charges partially cancels in their ratio. Near midrapidity in central collisions, the relative
SRC-induced enhancement is less pronounced than in
the 2.4 GeV default-mean-field comparison.
At 3.2 GeV, the SRC calculation remains close to the
STAR preliminary $\pi^-$ data, although it overpredicts
the $\pi^+$ central values.
A possible explanation is the reduced relative importance
of the initial intrinsic nucleon motion compared with
the larger beam-driven collision energy.
Because the mean-field parametrization also differs
between the two calculations, the present comparison
does not isolate a purely collision-energy-dependent effect.

\sect{Summary and discussion}

We have developed and implemented in SMASH a nuclear initialization scheme that jointly incorporates spatial nucleon correlations, high-momentum components, predominantly proton--neutron pair assignments, and correlated pair momenta.
With the default mean field, SRC initialization enhances both charged-pion yields at $\sqrt{s_{NN}}=2.4$~GeV, with a larger relative increase for $\pi^+$ and consequently a lower $\pi^-/\pi^+$ ratio. 

Controlled comparisons identify the initial high-momentum components as the dominant source of this enhancement. The tested spatial and pair-isospin variations have smaller effects, and enhanced production persists without the imposed pair structure. These results indicate that the charged-pion yields have limited ability to distinguish the finer correlation structures considered here. 
At $\sqrt{s_{NN}}=3.2$~GeV, calculations using the same momentum-independent mean field with $K\simeq240$~MeV for both initializations also show SRC-induced pion enhancement, with a smaller relative effect near midrapidity in central collisions.
The smaller relative enhancement at the higher collision energy may indicate a reduced sensitivity of pion production to the initial nucleon momenta.

Reaction histories show that the initial high-momentum components enhance resonance production. Stronger high-momentum repulsion acts in the opposite direction: it modifies nucleon propagation, lowers the mean kinetic invariant energy of accepted $NN$ collisions in the selected compression-stage window, and reduces resonance production, thereby counteracting the pion-yield enhancement. Using Gaussian-process emulators to guide the mean-field calibration against charged-pion multiplicities and proton elliptic flow, together with constraints on nuclear saturation properties, nuclear-matter pressure, and the nucleon effective mass, as well as stability and causality conditions, we obtain an incompressibility of $K\simeq215$~MeV, compared with $348.2$~MeV for the default parametrization. 

The effect of SRC-induced high-momentum tails on the high-density
EOS depends on the choice of mean-field model~\cite{Cai:2015xga,Cai:2022grw}.
In our transport calculations with SRC initialization, charged-pion
production is sensitive to the high-momentum dependence of the
nucleon mean field. At $\rho=0.160~\mathrm{fm}^{-3}$, the refitted
single-particle potential rises more steeply with kinetic energy
than the default potential and exceeds the Hama optical-model
reference at the upper end of the range shown in
Fig.~\ref{fig:optical}. This stronger repulsion is accompanied by
lower $\pi^+$ and $\pi^-$ yields. The short-range repulsive core
of the $NN$ interaction may contribute to the high-momentum
repulsion, although its contribution cannot be isolated by the
present fit.

Charged-kaon yields exhibit a similar qualitative response. The SRC calculation with the default mean field gives higher $K^+$ and $K^-$ yields than the WS reference, while the calculation with the refitted mean field gives lower yields. The charge ratios respond differently: SRC initialization lowers $\pi^-/\pi^+$ but raises $K^-/K^+$ in the default-mean-field calculations. The limited $K^-$ statistics do not allow a clear distinction between the $K^-/K^+$ ratios obtained with WS initialization and the default mean field and with SRC initialization and the refitted mean field.

The remaining underprediction of the charged-pion ratio
also motivates an assessment of the electromagnetic
interactions omitted here. Recent UrQMD calculations
for central Au+Au collisions at $\sqrt{s_{NN}}=2.42$ GeV
show that baryon--meson Coulomb interactions enhance
$\pi^-/\pi^+$ at low transverse momenta and suppress it
at higher momenta \cite{Li:2026Coulomb}.
Controlled transport comparisons for Sn+Sn collisions
at 270 MeV per nucleon also find an increase in the total
charged-pion ratio when Coulomb interactions are included
\cite{TMEP:2023ifw}.
These results make the omission of Coulomb dynamics
a plausible contributor to the remaining discrepancy.
Its magnitude, and whether it improves the present
$p_T$-integrated midrapidity ratio, must nevertheless
be determined for the same collision system, centrality
selection, and acceptance.

The joint sampling scheme retains explicit pair information
and provides a basis for identifying observables sensitive
to different SRC ingredients. Since the pion and kaon 
yields studied here respond primarily to the single-nucleon
momentum distribution, two-particle momentum correlations
and coincidence observables are natural candidates for
further tests of the implemented pair structure.

The interpretation of pion production also involves resonance potentials and in-medium reaction rates. Ikeno and Ono showed that repulsive contributions to the isoscalar $\Delta$ potential can reduce pion yields~\cite{Ikeno:2023cyc}, while recent UrQMD calculations examined the competing effects of in-medium $NN$ inelastic cross sections and SRC-induced high-momentum components~\cite{Li:2026dsx}. These studies identify additional dynamical ingredients relevant to the pion-production response. 
Studying these mechanisms with the present correlated
initialization would help disentangle initial-state
contributions from uncertainties in resonance dynamics
\cite{TMEP:2019yci,TMEP:2022xjg}.


\begin{acknowledgments}
This work is supported by the National Natural Science Foundation of China under No.\ 12435009, Grant No.\ 12075098. This work was supported by the Outstanding Leading Talent Team Program of Central China Normal University (XJ2026000302). The numerical calculation have been performed on the GPU cluster in the Nuclear Science Computing Center at Central China Normal University (NSC3). 
\end{acknowledgments}

\section*{Data Availability}
The data that support the findings of this article are openly available \cite{srcdataset}.

\appendix





\section{Time evolution of particle multiplicities and ratios}
\label{app:optical}

We examine the time evolution of the matched SRC-momentum and Fermi-momentum samples used in Table~\ref{table:reaction_table}, each containing 10,000 events. The samples share the same initial spatial configurations and proton/neutron labels and use the default mean field.

\begin{figure*}[htp]
\centering
\includegraphics[width=1\textwidth]{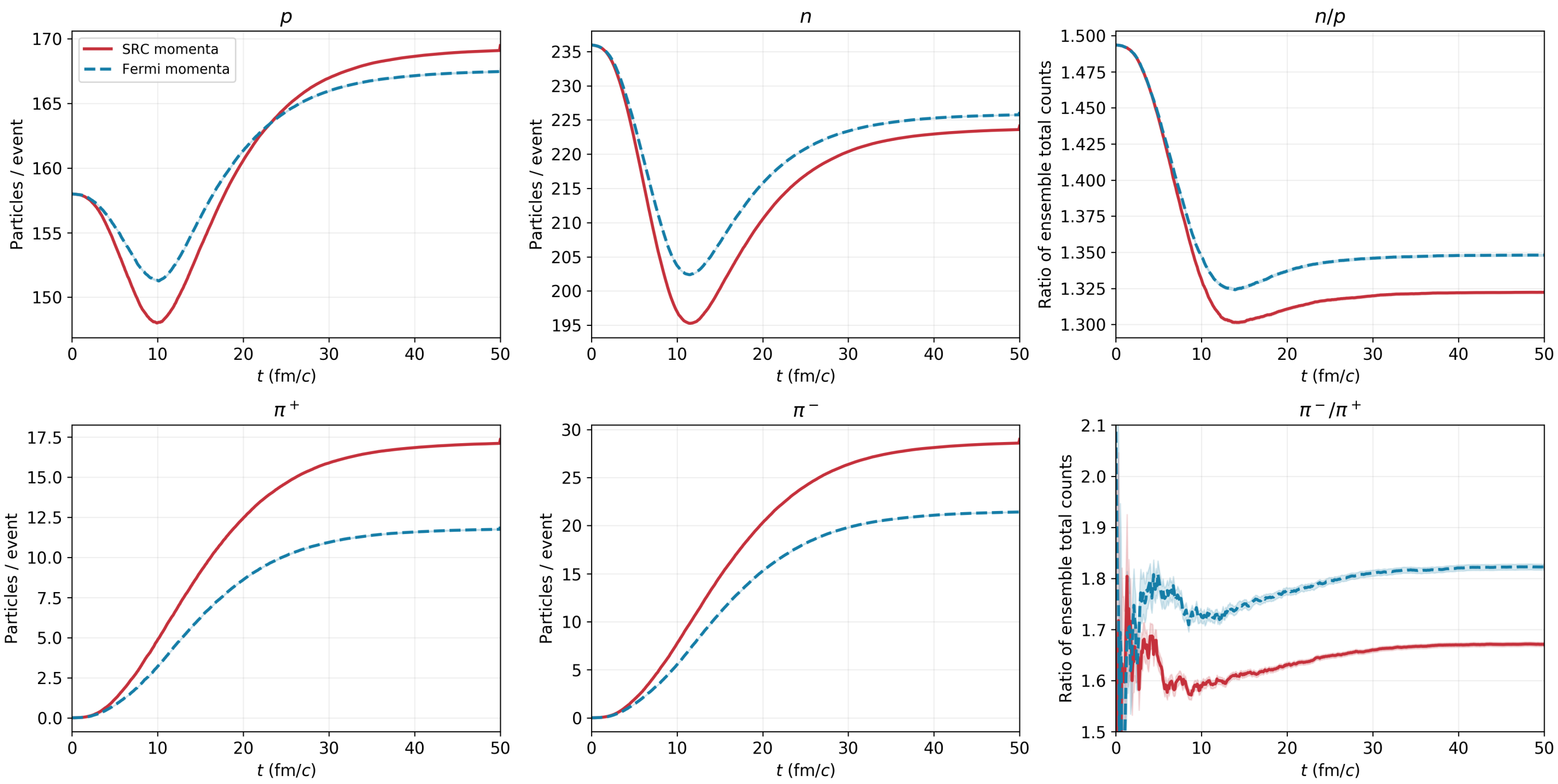}
\caption{(Color online)
Time evolution of the proton and neutron multiplicities,
$n/p$, charged-pion multiplicities, and $\pi^-/\pi^+$
in Au+Au collisions at $\sqrt{s_{NN}}=2.4$ GeV
and $b=0$.
}
\label{fig:b0_populations2p4}
\end{figure*}

Figure \ref{fig:b0_populations2p4} shows the time evolution of particle multiplicities and ratios for the matched samples B and C, which are also used in the reaction-channel analysis presented in Table \ref{table:reaction_table}. 
During the early evolution, the proton and neutron multiplicities initially decrease and subsequently recover. Because nucleons participate in resonance formation, these early changes cannot be interpreted directly as net neutron–proton conversion. At all recorded times with \(t>10~\mathrm{fm}/c\), both \(n/p\) and \(\pi^-/\pi^+\) are smaller in the SRC-momentum sample than in the Fermi-momentum sample.  The reduced pion ratio thus reflects a larger relative enhancement of \(\pi^+\), rather than suppression of \(\pi^-\).

\section{Mean-field calibration and screening criteria}
\label{app:mf_calibration}


We use Gaussian-process emulators to map
$\boldsymbol{\theta}=(K,C,\Lambda,S_{\mathrm{pot}})$
onto charged-pion multiplicities and differential
proton elliptic flow $v_2(p_T)$, and to guide
the selection of candidate parameters.

For each choice of $(K,C,\Lambda)$, the parameters
$(A,B,\tau)$ are determined by imposing the
saturation binding energy, zero pressure, and
specified incompressibility
\cite{Mohs:2024gyc}:
\begin{equation}
e(\rho_0)=-16~\mathrm{MeV},\qquad
P(\rho_0)=0,\qquad
9\rho_0^2 e''(\rho_0)=K,
\end{equation}
where $\rho_0=0.168~\mathrm{fm}^{-3}$. Here, $e(\rho)$ is the energy per nucleon excluding the rest mass, obtained by evaluating the kinetic and interaction energy densities for cold symmetric nuclear matter using the zero-temperature Fermi distribution and dividing their sum by $\rho$.
Unlike the calibration in Ref.~\cite{Mohs:2024gyc},
the present search does not impose optical-potential
constraints.

Additional screening uses the pressure, effective
$k$-mass, and sound speed of zero-temperature
symmetric nuclear matter.
The pressure is calculated as
\begin{equation}
P(\rho)=\rho^2\frac{de}{d\rho},
\end{equation}
where $e(\rho)$ is the energy per nucleon excluding
the rest mass.
We use the nonrelativistic effective $k$-mass
definition \cite{Cai:2022grw},
\begin{equation}
\frac{m_k^*}{m}
=
\left[
1+\frac{m}{p}\frac{\partial U}{\partial p}
\right]^{-1}_{\rho=\rho_0,\;p=p_F},
\end{equation}
evaluated in units with $c=1$.
The sound speed is obtained from
\begin{equation}
\frac{c_s^2}{c^2}=\frac{dP}{d\varepsilon},
\qquad
\varepsilon=\rho(mc^2+e).
\end{equation}

Candidate pressures are required to lie within
the flow-derived band for symmetric nuclear
matter of Ref.~\cite{Danielewicz:2002pu}.

We additionally adopt $0.5<m_k^*/m<0.85$ and
$10\leq S_{\mathrm{pot}}\leq20$ MeV as screening
intervals for the present search.
Candidates must also satisfy $0<c_s^2/c^2<1$
on the sampled density grid from $\rho_0$ to
$4\rho_0$.

We first retain parameter sets satisfying the physical screening conditions above. Gaussian-process emulators trained on SMASH calculations predict the charged-pion multiplicities and proton $v_2(p_T)$ for these candidates. Based on their weighted deviations from the experimental data, five candidate parameter sets are selected for direct SMASH calculations.

\section{Additional proton-flow observables}
\label{app:additional_flow}

We supplement the transverse-momentum dependence of proton
elliptic flow shown in Fig.~\ref{fig:v2pt2p4} with the rapidity
dependence of proton directed and elliptic flow. The calculations
use the same three ensembles at $\sqrt{s_{NN}}=2.4$ GeV and the
20--30\% centrality class: WS initialization with the default
mean field, SRC initialization with the default mean field,
and SRC initialization with the refitted mean field.

\begin{figure*}[htp]
\centering
\includegraphics[width=1\textwidth]{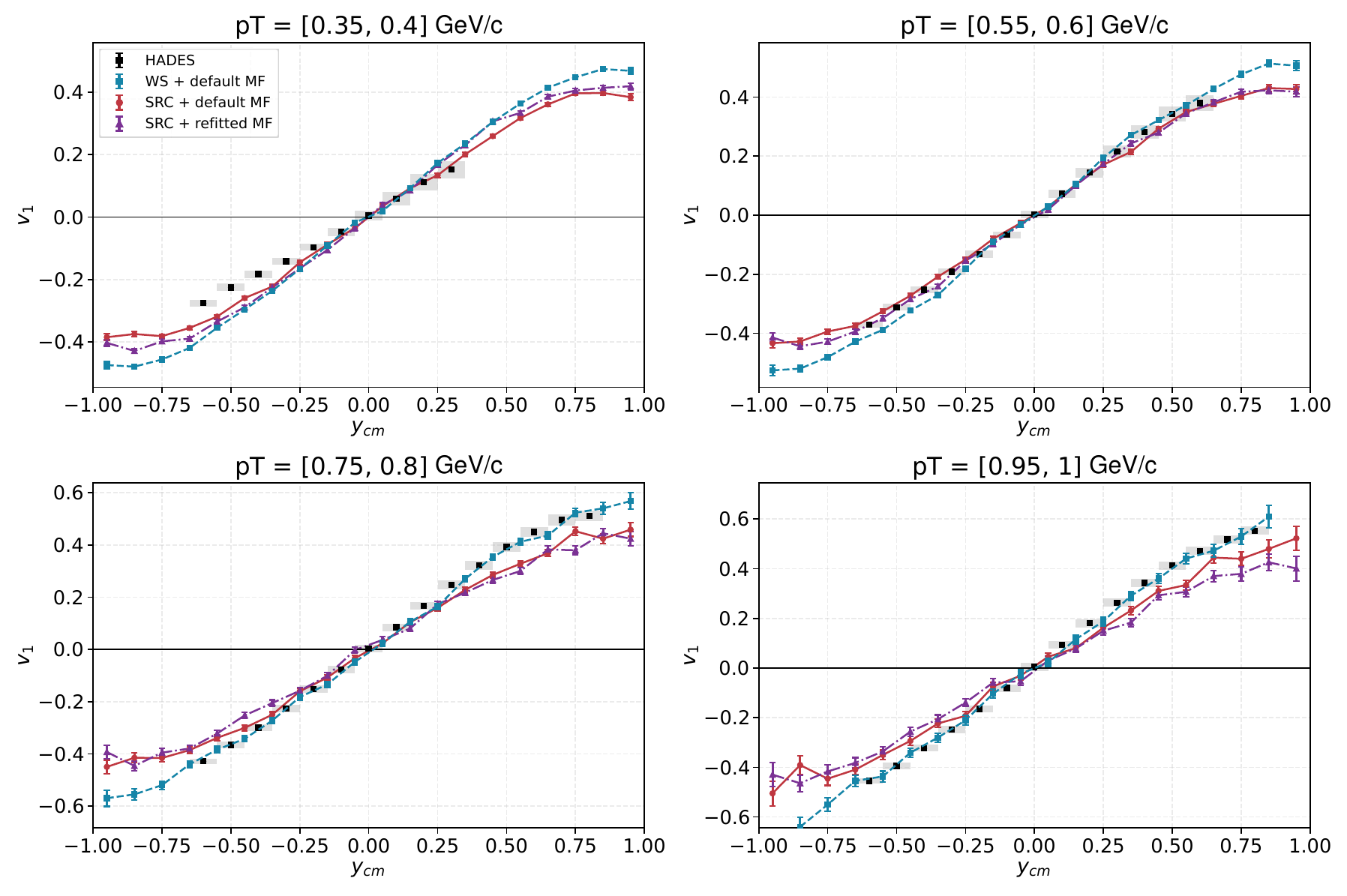}
\caption {\label{fig:v12p4}(Color online) Proton directed flow \(v_1\) as a function of center-of-mass rapidity \(y_{\mathrm{cm}}\) in 20–30\% central Au+Au collisions at \(\sqrt{s_{NN}}=2.4\) GeV, for the four transverse-momentum intervals indicated in the panels. SRC and WS results are compared with HADES measurements \cite{HADES:2022osk}.
}
\end{figure*}

Figure \ref{fig:v12p4} shows the proton directed flow \(v_1(y)\) in four transverse-momentum intervals for 20–30\% centrality. All three calculations cross zero near midrapidity, exhibiting positive and negative directed flow at positive and negative rapidities, respectively. In all four intervals, the curves are close near midrapidity and gradually separate with increasing absolute rapidity. At larger absolute rapidities within the range shown, the SRC calculations with the refitted and default mean fields exhibit similar trends, while the WS calculation with the default mean field generally yields a larger directed-flow magnitude than either SRC calculation.

\begin{figure*}[htp]
\centering
\includegraphics[width=1\textwidth]{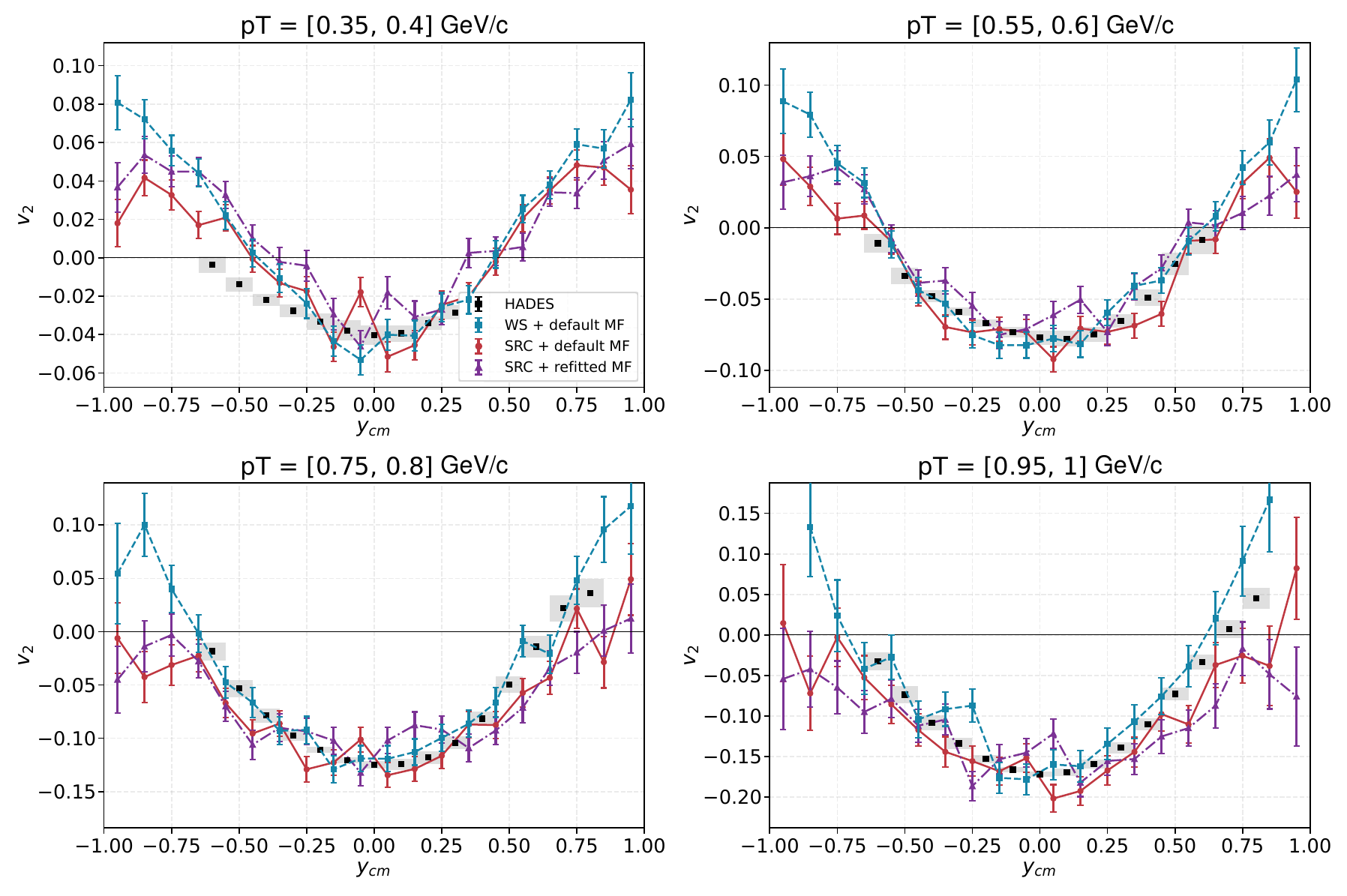}
\caption {\label{fig:v2y2p4}(Color online) Proton elliptic flow \(v_2\) as a function of center-of-mass rapidity in 20–30\% central Au+Au collisions at \(\sqrt{s_{NN}}=2.4\) GeV, for the four transverse-momentum intervals indicated in the panels. SRC and WS results are compared with HADES measurements \cite{HADES:2022osk}.
}
\end{figure*}

Figure \ref{fig:v2y2p4} shows the rapidity dependence of elliptic flow in different transverse-momentum intervals. All calculations exhibit negative elliptic flow near midrapidity, with \(v_2\) increasing as the absolute rapidity increases. Higher transverse-momentum intervals exhibit a larger magnitude of negative elliptic flow near midrapidity.
Together, Figs.~\ref{fig:v12p4} and \ref{fig:v2y2p4}
show that the initialization and mean-field adjustment
modify the magnitude and kinematic dependence of proton
flow while retaining its overall rapidity dependence.

\section{Charged-pion transverse-mass spectra}
\label{app:pion_mt}

We supplement the rapidity distributions in the main text
with reduced transverse-mass spectra for the same three
calculations.

\begin{figure*}[htp]
\centering
\includegraphics[width=\textwidth]{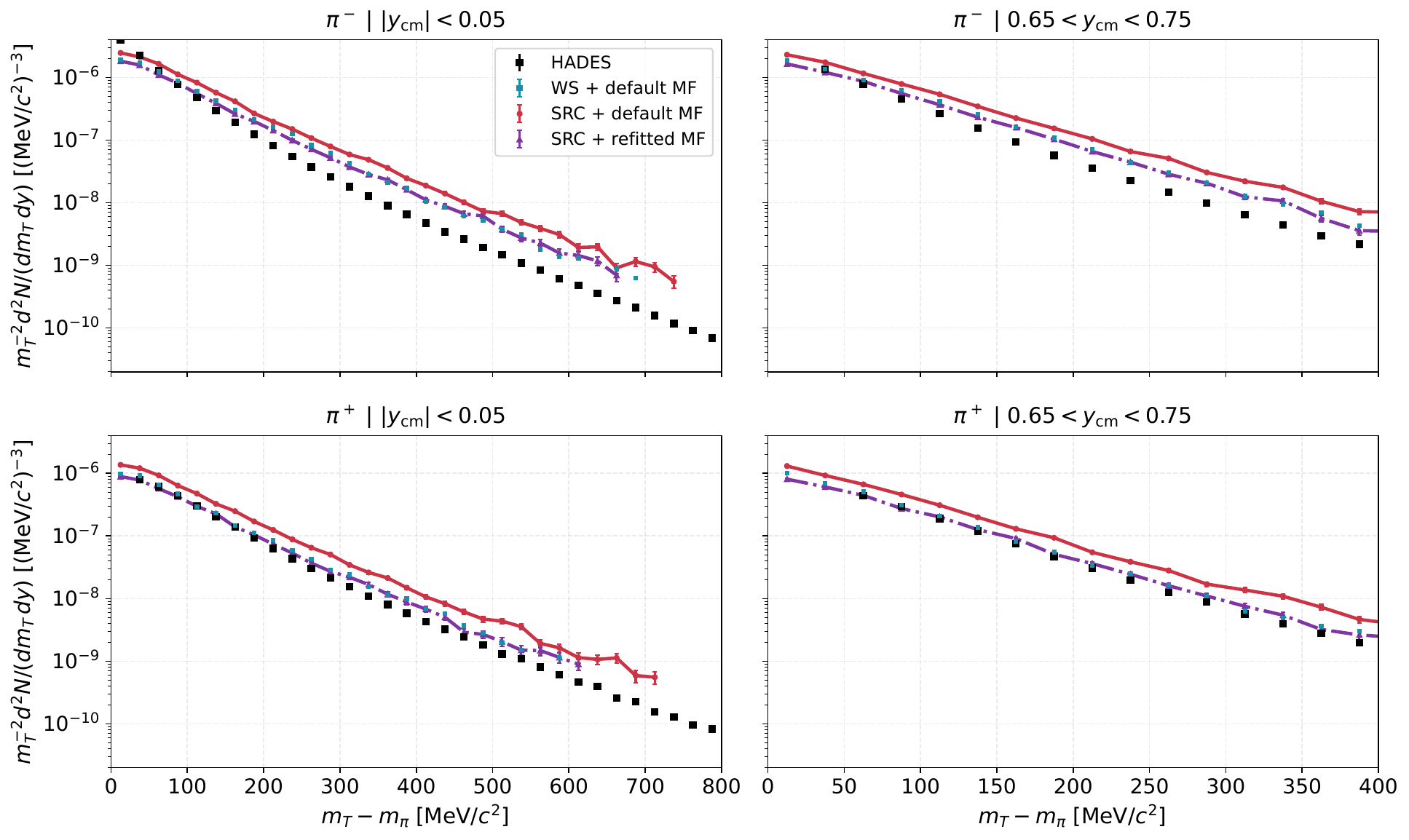}
\caption{(Color online)
Reduced transverse-mass spectra,
$m_T^{-2}d^2N/(dm_T\,dy)$, of $\pi^-$ (upper panels)
and $\pi^+$ (lower panels) in 0--10\% central Au+Au
collisions at $\sqrt{s_{NN}}=2.4$ GeV, plotted as a
function of $m_T-m_\pi$.
The left and right columns correspond to
$|y_{\mathrm{cm}}|<0.05$ and
$0.65<y_{\mathrm{cm}}<0.75$, respectively.
Black squares show the HADES data from Fig.~5 of
Ref.~\cite{HADES:2020ver}.
}
\label{fig:mt2p4}
\end{figure*}

Figure~\ref{fig:mt2p4} shows the reduced transverse-mass
spectra, $m_T^{-2}d^2N/(dm_T\,dy)$, of charged pions in
0--10\% central Au+Au collisions at
$\sqrt{s_{NN}}=2.4$ GeV, plotted against $m_T-m_\pi$.
Here, $m_T=\sqrt{m_\pi^2+p_T^2}$.
The two columns correspond to midrapidity and forward
rapidity, respectively.

With the default mean field, the SRC calculation yields
larger spectral differential yields than the WS reference over
most of the displayed range. The SRC calculation with
the refitted mean field gives lower differential yields and
generally lies closer to the WS result.
Within the displayed forward-rapidity range, the refitted
calculation agrees more closely with the $\pi^+$ data
than with the $\pi^-$ data.
At midrapidity, appreciable excesses remain at intermediate
and large $m_T-m_\pi$, particularly for $\pi^-$.

\section{Variations of incompressibility and
symmetry-potential strength}
\label{app:KS_scan}

To test whether mean-field adjustments can compensate
for the SRC-induced pion enhancement, we first vary
the incompressibility $K$ and the symmetry-potential
strength $S_{\rm pot}$ while keeping $C_{\rm MD}$
and $\Lambda$ fixed. 
For each $K$, the parameters $A$, $B$, and $\tau$ were recalibrated
to preserve the saturation energy and vanishing pressure at saturation.
As shown in Table~\ref{tab:KS_pion_scan}, increasing $K$
at fixed momentum-dependent parameters reduces the
charged-pion multiplicities over part of the explored
range, but the overall effect is limited and does not
remove their excess over the experimental central values.
Increasing $S_{\mathrm{pot}}$ slightly raises the $\pi^+$ multiplicity
and lowers the $\pi^-$ multiplicity, without bringing both into
agreement with the data, as shown in \ref{tab:KS_pion_scan}.
Within the explored parameter range and the present transport setup,
adjusting $K$ and $S_{\mathrm{pot}}$ alone is therefore insufficient
to compensate for the SRC-induced enhancement.

\begin{table}[!htb]
\centering
\caption{Charged-pion multiplicities obtained by varying $K$
and $S_{\mathrm{pot}}$ at fixed momentum-dependent parameters
$C_{\mathrm{MD}}=-63.1052$ MeV and
$\Lambda=2.11952$ fm$^{-1}$.
All calculations employ SRC initial states for Au+Au collisions
at $\sqrt{s_{NN}}=2.4$ GeV in the 20--30\% centrality class.
The experimental central values are shown for comparison.}
\label{tab:KS_pion_scan}
\begin{tabular}{cccc}
\hline\hline
$K$ (MeV) & $S_{\mathrm{pot}}$ (MeV)
& $\langle N_{\pi^+}\rangle$
& $\langle N_{\pi^-}\rangle$ \\
\hline
348.2 & 15.8 & 7.733 & 13.488 \\
500   & 15.8 & 7.281 & 12.807 \\
600   & 15.8 & 7.127 & 12.652 \\
348.2 & 60   & 7.867 & 13.269 \\
500   & 60   & 7.506 & 12.440 \\
\hline
\multicolumn{2}{c}{Experiment}
& 4.7 & 8.7 \\
\hline\hline
\end{tabular}
\end{table}

\bibliography{ref}

@article{HADES:2020ver,
    author = "Adamczewski-Musch, J. and others",
    collaboration = "HADES",
    title = "{Charged-pion production in $\mathbf {Au+Au}$ collisions at $\sqrt{\mathbf {s}_{\mathbf {NN}}} = 2.4~{\mathbf {GeV}}$: HADES Collaboration}",
    eprint = "2005.08774",
    archivePrefix = "arXiv",
    primaryClass = "nucl-ex",
    doi = "10.1140/epja/s10050-020-00237-2",
    journal = "Eur. Phys. J. A",
    volume = "56",
    number = "10",
    pages = "259",
    year = "2020"
}

@article{HADES:2022osk,
    author = "Adamczewski-Musch, J. and others",
    collaboration = "HADES",
    title = "{Proton, deuteron and triton flow measurements in Au+Au collisions at $\sqrt{s_{_{{\text {NN}}}}}= 2.4$~GeV}",
    eprint = "2208.02740",
    archivePrefix = "arXiv",
    primaryClass = "nucl-ex",
    doi = "10.1140/epja/s10050-023-00936-6",
    journal = "Eur. Phys. J. A",
    volume = "59",
    number = "4",
    pages = "80",
    year = "2023"
}

@article{Subedi:2008zz,
    author = "Subedi, R. and others",
    title = "{Probing Cold Dense Nuclear Matter}",
    eprint = "0908.1514",
    archivePrefix = "arXiv",
    primaryClass = "nucl-ex",
    reportNumber = "JLAB-PHY-08-828",
    doi = "10.1126/science.1156675",
    journal = "Science",
    volume = "320",
    pages = "1476--1478",
    year = "2008"
}

@article{CLAS:2018xvc,
    author = "Duer, M. and others",
    collaboration = "CLAS",
    title = "{Direct Observation of Proton-Neutron Short-Range Correlation Dominance in Heavy Nuclei}",
    eprint = "1810.05343",
    archivePrefix = "arXiv",
    primaryClass = "nucl-ex",
    doi = "10.1103/PhysRevLett.122.172502",
    journal = "Phys. Rev. Lett.",
    volume = "122",
    number = "17",
    pages = "172502",
    year = "2019"
}

@article{CLAS:2018qpc,
    author = "Cohen, E. O. and others",
    collaboration = "CLAS",
    title = "{Center of Mass Motion of Short-Range Correlated Nucleon Pairs studied via the $A(e,e^{\prime}pp)$ Reaction}",
    eprint = "1805.01981",
    archivePrefix = "arXiv",
    primaryClass = "nucl-ex",
    doi = "10.1103/PhysRevLett.121.092501",
    journal = "Phys. Rev. Lett.",
    volume = "121",
    number = "9",
    pages = "092501",
    year = "2018"
}

@article{Alvioli:2016wwp,
    author = "Alvioli, Massimiliano and Ciofi degli Atti, Claudio and Morita, Hiko",
    title = "{Universality of nucleon-nucleon short-range correlations: the factorization property of the nuclear wave function, the relative and center-of-mass momentum distributions, and the nuclear contacts}",
    eprint = "1607.04103",
    archivePrefix = "arXiv",
    primaryClass = "nucl-th",
    doi = "10.1103/PhysRevC.94.044309",
    journal = "Phys. Rev. C",
    volume = "94",
    number = "4",
    pages = "044309",
    year = "2016"
}

@article{Yang:2018xtl,
    author = "Yang, Zu-Xing and Fan, Xiao-Hua and Yong, Gao-Chan and Zuo, Wei",
    title = "{Effects of the initialization of nucleon momentum in heavy-ion collisions at medium energies}",
    eprint = "1807.06727",
    archivePrefix = "arXiv",
    primaryClass = "nucl-th",
    doi = "10.1103/PhysRevC.98.014623",
    journal = "Phys. Rev. C",
    volume = "98",
    number = "1",
    pages = "014623",
    year = "2018"
}

@misc{Li:2026dsx,
    author = "Li, Pengcheng and Larionov, A. B. and Wang, Yongjia and Yong, Gaochan and Li, Qingfeng",
    title = "{Effects of in-medium $NN$ inelastic cross sections and the high-momentum tail of nucleon momentum distributions on pion production in heavy-ion collisions}",
    eprint = "2609.25818",
    archivePrefix = "arXiv",
    primaryClass = "nucl-th",
    month = "9",
    year = "2026"
}

@article{TMEP:2023ifw,
    author = "Xu, Jun and others",
    collaboration = "TMEP",
    title = "{Comparing pion production in transport simulations of heavy-ion collisions at 270AMeV under controlled conditions}",
    eprint = "2308.05347",
    archivePrefix = "arXiv",
    primaryClass = "nucl-th",
    doi = "10.1103/PhysRevC.109.044609",
    journal = "Phys. Rev. C",
    volume = "109",
    number = "4",
    pages = "044609",
    year = "2024"
}

@article{SMASH:2016zqf,
    author = "Weil, J. and others",
    collaboration = "SMASH",
    title = "{Particle production and equilibrium properties within a new hadron transport approach for heavy-ion collisions}",
    eprint = "1606.06642",
    archivePrefix = "arXiv",
    primaryClass = "nucl-th",
    doi = "10.1103/PhysRevC.94.054905",
    journal = "Phys. Rev. C",
    volume = "94",
    number = "5",
    pages = "054905",
    year = "2016"
}

@misc{Liu:2026tde,
    author = "Liu, Lu-Meng and Xu, Jun and Huang, Xu-Guang",
    title = "{Measuring short-range correlations using relativistic heavy-ion collisions}",
    eprint = "2609.29237",
    archivePrefix = "arXiv",
    primaryClass = "nucl-th",
    month = "9",
    year = "2026"
}

@article{Mohs:2024gyc,
    author = "Mohs, Justin and Spies, Simon and Elfner, Hannah",
    title = "{Constraints on the equation~of state of nuclear matter from systematically comparing SMASH calculations to HADES data}",
    eprint = "2409.16927",
    archivePrefix = "arXiv",
    primaryClass = "nucl-th",
    doi = "10.1103/f9zw-ygwh",
    journal = "Phys. Rev. C",
    volume = "112",
    number = "4",
    pages = "044905",
    year = "2025"
}

@article{Cai:2015xga,
    author = "Cai, Bao-Jun and Li, Bao-An",
    title = "{Symmetry energy of cold nucleonic matter within a relativistic mean field model encapsulating effects of high momentum nucleons induced by short-range correlations}",
    eprint = "1509.09290",
    archivePrefix = "arXiv",
    primaryClass = "nucl-th",
    doi = "10.1103/PhysRevC.93.014619",
    journal = "Phys. Rev. C",
    volume = "93",
    number = "1",
    pages = "014619",
    year = "2016"
}

@article{HADES:2022mwn,
    author = "Adamczewski-Musch, J. and others",
    collaboration = "HADES",
    title = "{Impact of the Coulomb field on charged-pion spectra in few-GeV heavy-ion collisions}",
    eprint = "2202.12750",
    archivePrefix = "arXiv",
    primaryClass = "nucl-ex",
    doi = "10.1140/epja/s10050-022-00796-6",
    journal = "Eur. Phys. J. A",
    volume = "58",
    number = "9",
    pages = "166",
    year = "2022"
}

@misc{Li:2025bdn,
    author = "Li, Pei and Sun, Kai-Jia and Zhou, Bo and Ma, Guo-Liang",
    title = "{Universal imprinting of short-range correlations in relativistic heavy-ion collisions}",
    eprint = "2511.23293",
    archivePrefix = "arXiv",
    primaryClass = "nucl-th",
    month = "11",
    year = "2025"
}

@misc{Chen_STAR_CPOD2026,
  author       = {Chen, Liubing},
  title        = {Charged Pion Production in {Au+Au} Collisions at
                  $\sqrt{s_{NN}}=3.0$--$4.5$ {GeV}
                  with the {STAR} Detector},
  year         = {2026},
  howpublished = {Presentation at CPOD 2026, CERN,
                  Geneva, Switzerland, 13--17 April 2026},
  note         = {For the STAR Collaboration},
  url          = {https://indico.cern.ch/event/1438306/contributions/6885850/}
}

@misc{Cai:2022grw,
    author = "Cai, Bao-Jun and Li, Bao-An",
    title = "{Nuclear Equation of State and Single-nucleon Potential from Gogny-like Energy Density Functionals Encapsulating Effects of Nucleon-nucleon Short-range Correlations}",
    eprint = "2210.10924",
    archivePrefix = "arXiv",
    primaryClass = "nucl-th",
    month = "10",
    year = "2022"
}

@article{Hama:1990vr,
    author = "Hama, S. and Clark, B. C. and Cooper, E. D. and Sherif, H. S. and Mercer, R. L.",
    title = "{Global Dirac optical potentials for elastic proton scattering from heavy nuclei}",
    doi = "10.1103/PhysRevC.41.2737",
    journal = "Phys. Rev. C",
    volume = "41",
    pages = "2737--2755",
    year = "1990"
}

@article{Alvioli:2005cz,
    author = "Alvioli, M. and Ciofi degli Atti, C. and Morita, H.",
    title = "{Ground-state energies, densities and momentum distributions in closed-shell nuclei calculated within a cluster expansion approach and realistic interactions}",
    doi = "10.1103/PhysRevC.72.054310",
    journal = "Phys. Rev. C",
    volume = "72",
    pages = "054310",
    year = "2005"
}

@article{Alvioli:2009ab,
    author = "Alvioli, M. and Drescher, H. -J. and Strikman, M.",
    title = "{A Monte Carlo generator of nucleon configurations in complex nuclei including nucleon-nucleon correlations}",
    doi = "10.1016/j.physletb.2009.08.067",
    journal = "Phys. Lett. B",
    volume = "680",
    pages = "225--230",
    year = "2009"
}

@article{Welke:1988zz,
    author = "Welke, G. M. and Prakash, M. and Kuo, T. T. S. and Das Gupta, S. and Gale, Charles",
    title = "{Azimuthal distributions in heavy ion collisions and the nuclear equation of state}",
    doi = "10.1103/PhysRevC.38.2101",
    journal = "Phys. Rev. C",
    volume = "38",
    pages = "2101--2107",
    year = "1988"
}

@article{Tarasovicova:2024isp,
    author = "Tarasovi{\v{c}}ov{\'a}, Lucia Anna and Mohs, Justin and Andronic, Anton and Elfner, Hannah and Kampert, Karl-Heinz",
    title = "{Flow and equation of state of nuclear matter at $E_{\mathrm{kin}}/A=0.25${\textendash}1.5~GeV with the SMASH transport approach}",
    eprint = "2405.09889",
    archivePrefix = "arXiv",
    primaryClass = "nucl-th",
    doi = "10.1140/epja/s10050-024-01445-w",
    journal = "Eur. Phys. J. A",
    volume = "60",
    number = "11",
    pages = "232",
    year = "2024"
}

@article{Danielewicz:2002pu,
    author = "Danielewicz, Pawel and Lacey, Roy and Lynch, William G.",
    title = "{Determination of the equation of state of dense matter}",
    eprint = "nucl-th/0208016",
    archivePrefix = "arXiv",
    doi = "10.1126/science.1078070",
    journal = "Science",
    volume = "298",
    pages = "1592--1596",
    year = "2002"
}

@article{Ikeno:2023cyc,
  author        = {Ikeno, Natsumi and Ono, Akira},
  title         = {Collision integral with momentum-dependent
                   potentials and its impact on pion production
                   in heavy-ion collisions},
  journal       = {Phys. Rev. C},
  volume        = {108},
  pages         = {044601},
  year          = {2023},
  doi           = {10.1103/PhysRevC.108.044601},
  eprint        = {2307.02395},
  archivePrefix = {arXiv},
  primaryClass  = {nucl-th}
}

@article{Egiyan2006SRCProbabilities,
  author = {Egiyan, K. S. and others},
  collaboration = {CLAS},
  title = {Measurement of 2- and 3-Nucleon Short Range Correlation Probabilities in Nuclei},
  journal = {Phys. Rev. Lett.},
  volume = {96},
  pages = {082501},
  year = {2006},
  doi = {10.1103/PhysRevLett.96.082501},
  eprint = {nucl-ex/0508026},
  archivePrefix = {arXiv}
}

@article{Li:2026Coulomb,
  author = {Li, Pengcheng and Wang, Yongjia and Li, Qingfeng},
  title = {Dynamical Coulomb effects on charged-pion
           HBT-radius splitting in central Au+Au collisions
           at few-GeV energies},
  journal = {Phys. Lett. B},
  volume = {882},
  pages = {141081},
  year = {2026},
  doi = {10.1016/j.physletb.2026.141081},
  eprint = {2609.23422},
  archivePrefix = {arXiv},
  primaryClass = {nucl-th}
}

@misc{Li2025MeanFieldPions,
  author = {Li, Xin and Wang, Si-Pei and Zhang, Zhen
            and Wang, Rui and Pu, Jie and Ma, Chun-Wang
            and Chen, Lie-Wen},
  title = {Effects of mean-field momentum dependence
           on pion production in intermediate-energy
           heavy-ion collisions},
  eprint = {2509.21099},
  archivePrefix = {arXiv},
  primaryClass = {nucl-th},
  year = {2025}
}

@article{Bisconti2007Momentum,
  author        = {Bisconti, C. and Arias de Saavedra, F. and Co', G.},
  title         = {Momentum distributions and spectroscopic factors
                   of doubly-closed shell nuclei in correlated
                   basis function theory},
  journal       = {Phys. Rev. C},
  volume        = {75},
  pages         = {054302},
  year          = {2007},
  doi           = {10.1103/PhysRevC.75.054302},
  eprint        = {nucl-th/0702061},
  archivePrefix = {arXiv}
}

@article{Ciofi2015Review,
  author = {{Ciofi degli Atti}, Claudio},
  title = {{In-medium short-range dynamics of nucleons: Recent theoretical and experimental advances}},
  journal = {Phys. Rept.},
  volume = {590},
  pages = {1--85},
  year = {2015},
  doi = {10.1016/j.physrep.2015.06.002}
}

@article{Hen2017Review,
  author = {Hen, Or and Miller, Gerald A. and Piasetzky, Eli and Weinstein, Lawrence B.},
  title = {{Nucleon-nucleon correlations, short-lived excitations, and the quarks within}},
  journal = {Rev. Mod. Phys.},
  volume = {89},
  number = {4},
  pages = {045002},
  year = {2017},
  doi = {10.1103/RevModPhys.89.045002},
  eprint = {1611.09748},
  archivePrefix = {arXiv},
  primaryClass = {nucl-ex}
}

@article{Wang2025Ultracentral,
  author = {Wang, Qi and Pang, Long-Gang and Wang, Xin-Nian},
  title = {{Addressing the Ultra-Central Puzzle with Initial-State Nuclear Structures}},
  journal = {Chin. Phys. Lett.},
  volume = {43},
  number = {2},
  pages = {020101},
  year = {2026},
  doi = {10.1088/0256-307X/43/2/020101},
  eprint = {2504.19208},
  archivePrefix = {arXiv},
  primaryClass = {nucl-th}
}

@misc{Liu2025IsolatedNN,
  author = {Liu, Qi and Mehrabpour, Hadi and Lu, Bing-Nan},
  title = {{Impacts of isolated nucleon-nucleon correlations in relativistic $^{16}$O+$^{16}$O collisions}},
  year = {2025},
  eprint = {2509.00315},
  archivePrefix = {arXiv},
  primaryClass = {nucl-th},
  url = {https://arxiv.org/abs/2509.00315v2}
}

@misc{Liu2026OOSRC,
  author = {Liu, Qi and Wu, Shanjin and Song, Huichao},
  title = {{Exploring the short-range correlations in $^{16}$O+$^{16}$O collisions at the LHC}},
  year = {2026},
  eprint = {2609.27566},
  archivePrefix = {arXiv},
  primaryClass = {nucl-th},
  url = {https://arxiv.org/abs/2609.27566}
}

@article{Huang2025NNCorrelations,
  author = {Huang, Yu-Jing and Meng, Zhu and Pang, Long-Gang and Wang, Xin-Nian},
  title = {{Novel deep learning method for detecting nucleon-nucleon correlations}},
  journal = {Phys. Rev. C},
  volume = {113},
  number = {3},
  pages = {034916},
  year = {2026},
  doi = {10.1103/hymw-tknk},
  eprint = {2504.00790},
  archivePrefix = {arXiv},
  primaryClass = {nucl-th}
}

@article{HADES:2017jgz,
    author = "Adamczewski-Musch, J. and others",
    collaboration = "HADES",
    title = "{Deep sub-threshold $\phi$ production in Au+Au collisions}",
    eprint = "1703.08418",
    archivePrefix = "arXiv",
    primaryClass = "nucl-ex",
    doi = "10.1016/j.physletb.2018.01.048",
    journal = "Phys. Lett. B",
    volume = "778",
    pages = "403--407",
    year = "2018"
}

@phdthesis{Schuldes2016Kaons,
  author = {Schuldes, Heidi},
  title = {Charged Kaon and {$\phi$} Reconstruction
           in {Au+Au} Collisions at 1.23 {AGeV}},
  school = {Johann Wolfgang Goethe-Universit{\"a}t
            Frankfurt am Main},
  year = {2016},
  url = {https://nbn-resolving.org/urn:nbn:de:hebis:30:3-424894}
}

@misc{Denicol:2014ywa,
    author = "Denicol, G. S. and Gale, C. and Jeon, S. and Paquet, J. -F. and Schenke, B.",
    title = "{Effect of initial-state nucleon-nucleon correlations on collective flow in ultra-central heavy-ion collisions}",
    eprint = "1406.7792",
    archivePrefix = "arXiv",
    primaryClass = "nucl-th",
    month = "6",
    year = "2014"
}

@article{Cruz-Torres:2019fum,
    author = "Cruz-Torres, R. and Lonardoni, D. and Weiss, R. and Barnea, N. and Higinbotham, D. W. and Piasetzky, E. and Schmidt, A. and Weinstein, L. B. and Wiringa, R. B. and Hen, O.",
    title = "{Many-body factorization and position{\textendash}momentum equivalence of nuclear short-range correlations}",
    eprint = "1907.03658",
    archivePrefix = "arXiv",
    primaryClass = "nucl-th",
    reportNumber = "LA-UR-19-25832",
    doi = "10.1038/s41567-020-01053-7",
    journal = "Nature Phys.",
    volume = "17",
    number = "3",
    pages = "306--310",
    year = "2021"
}

@article{Weiss:2015mba,
    author = "Weiss, Ronen and Bazak, Betzalel and Barnea, Nir",
    title = "{Generalized nuclear contacts and momentum distributions}",
    eprint = "1503.07047",
    archivePrefix = "arXiv",
    primaryClass = "nucl-th",
    doi = "10.1103/PhysRevC.92.054311",
    journal = "Phys. Rev. C",
    volume = "92",
    number = "5",
    pages = "054311",
    year = "2015"
}

@article{Wiringa:2013ala,
    author = "Wiringa, R. B. and Schiavilla, R. and Pieper, Steven C. and Carlson, J.",
    title = "{Nucleon and nucleon-pair momentum distributions in $A \le 12$ nuclei}",
    eprint = "1309.3794",
    archivePrefix = "arXiv",
    primaryClass = "nucl-th",
    reportNumber = "JLAB-THY-13-1789",
    doi = "10.1103/PhysRevC.89.024305",
    journal = "Phys. Rev. C",
    volume = "89",
    number = "2",
    pages = "024305",
    year = "2014"
}

@article{Casale:2023ews,
    author = "Casale, P. R. and Amaro, J. E. and Ruiz Arriola, E. and Ruiz Simo, I.",
    title = "{Center-of-mass momentum dependence of short-range correlations with the coarse-grained Granada potential}",
    eprint = "2304.00296",
    archivePrefix = "arXiv",
    primaryClass = "nucl-th",
    doi = "10.1103/PhysRevC.108.054001",
    journal = "Phys. Rev. C",
    volume = "108",
    number = "5",
    pages = "054001",
    year = "2023"
}

@article{TMEP:2022xjg,
    author = "Wolter, Hermann and others",
    collaboration = "TMEP",
    title = "{Transport model comparison studies of intermediate-energy heavy-ion collisions}",
    eprint = "2202.06672",
    archivePrefix = "arXiv",
    primaryClass = "nucl-th",
    doi = "10.1016/j.ppnp.2022.103962",
    journal = "Prog. Part. Nucl. Phys.",
    volume = "125",
    pages = "103962",
    year = "2022"
}

@article{Hong:2013yva,
    author = "Hong, Jun and Danielewicz, P.",
    title = "{Subthreshold pion production within a transport description of central Au + Au collisions}",
    eprint = "1307.7654",
    archivePrefix = "arXiv",
    primaryClass = "nucl-th",
    doi = "10.1103/PhysRevC.90.024605",
    journal = "Phys. Rev. C",
    volume = "90",
    number = "2",
    pages = "024605",
    year = "2014"
}

@article{TMEP:2019yci,
    author = "Ono, Akira and others",
    collaboration = "TMEP",
    title = "{Comparison of heavy-ion transport simulations: Collision integral with pions and {\ensuremath{\Delta}} resonances in a box}",
    eprint = "1904.02888",
    archivePrefix = "arXiv",
    primaryClass = "nucl-th",
    doi = "10.1103/PhysRevC.100.044617",
    journal = "Phys. Rev. C",
    volume = "100",
    number = "4",
    pages = "044617",
    year = "2019"
}

@article{Danielewicz:1999zn,
    author = "Danielewicz, Pawel",
    title = "{Determination of the mean field momentum dependence using elliptic flow}",
    eprint = "nucl-th/9912027",
    archivePrefix = "arXiv",
    reportNumber = "MSUCL-1143",
    doi = "10.1016/S0375-9474(00)00083-X",
    journal = "Nucl. Phys. A",
    volume = "673",
    pages = "375--410",
    year = "2000"
}

@article{CLAS:2018yvt,
    author = "Duer, M. and others",
    collaboration = "CLAS",
    title = "{Probing high-momentum protons and neutrons in neutron-rich nuclei}",
    doi = "10.1038/s41586-018-0400-z",
    journal = "Nature",
    volume = "560",
    number = "7720",
    pages = "617--621",
    year = "2018"
}

@article{Hen:2014nza,
    author = "Hen, O. and others",
    title = "{Momentum sharing in imbalanced Fermi systems}",
    eprint = "1412.0138",
    archivePrefix = "arXiv",
    primaryClass = "nucl-ex",
    doi = "10.1126/science.1256785",
    journal = "Science",
    volume = "346",
    pages = "614--617",
    year = "2014"
}

@article{Steinberg:2018jvv,
    author = {Steinberg, V. and Staudenmaier, J. and Oliinychenko, D. and Li, F. and Erkiner, {\"O}. and Elfner, H.},
    collaboration = "SMASH",
    title = "{Strangeness production via resonances in heavy-ion collisions at energies available at the GSI Schwerionensynchrotron}",
    eprint = "1809.03828",
    archivePrefix = "arXiv",
    primaryClass = "nucl-th",
    doi = "10.1103/PhysRevC.99.064908",
    journal = "Phys. Rev. C",
    volume = "99",
    number = "6",
    pages = "064908",
    year = "2019"
}

@article{CLAS:2020rue,
    author = "Korover, I. and others",
    collaboration = "CLAS",
    title = "{12C(e,e'pN) measurements of short range correlations in the tensor-to-scalar interaction transition region}",
    eprint = "2004.07304",
    archivePrefix = "arXiv",
    primaryClass = "nucl-ex",
    doi = "10.1016/j.physletb.2021.136523",
    journal = "Phys. Lett. B",
    volume = "820",
    pages = "136523",
    year = "2021"
}

@article{CLAS:2020mom,
    author = "Schmidt, A. and others",
    collaboration = "CLAS",
    title = "{Probing the core of the strong nuclear interaction}",
    eprint = "2004.11221",
    archivePrefix = "arXiv",
    primaryClass = "nucl-ex",
    doi = "10.1038/s41586-020-2021-6",
    journal = "Nature",
    volume = "578",
    number = "7796",
    pages = "540--544",
    year = "2020"
}

@article{Arrington:2022sov,
    author = "Arrington, John and Fomin, Nadia and Schmidt, Axel",
    title = "{Progress in understanding short-range structure in nuclei: an experimental perspective}",
    eprint = "2203.02608",
    archivePrefix = "arXiv",
    primaryClass = "nucl-ex",
    doi = "10.1146/annurev-nucl-102020-022253",
    journal = "Ann. Rev. Nucl. Part. Sci.",
    volume = "72",
    pages = "307--337",
    year = "2022"
}

@article{Rios:2013zqa,
    author = "Rios, A. and Polls, A. and Dickhoff, W. H.",
    title = "{Density and isospin asymmetry dependence of high-momentum components}",
    eprint = "1312.7307",
    archivePrefix = "arXiv",
    primaryClass = "nucl-th",
    doi = "10.1103/PhysRevC.89.044303",
    journal = "Phys. Rev. C",
    volume = "89",
    number = "4",
    pages = "044303",
    year = "2014"
}

@article{Ryckebusch:2014ann,
    author = "Ryckebusch, Jan and Cosyn, Wim and Vanhalst, Maarten",
    title = "{Stylized features of single-nucleon momentum distributions}",
    eprint = "1405.3814",
    archivePrefix = "arXiv",
    primaryClass = "nucl-th",
    doi = "10.1088/0954-3899/42/5/055104",
    journal = "J. Phys. G",
    volume = "42",
    number = "5",
    pages = "055104",
    year = "2015"
}

@article{Lonardoni:2018sqo,
    author = "Lonardoni, D. and Gandolfi, S. and Wang, X. B. and Carlson, J.",
    title = "{Single- and two-nucleon momentum distributions for local chiral interactions}",
    eprint = "1804.08027",
    archivePrefix = "arXiv",
    primaryClass = "nucl-th",
    reportNumber = "LA-UR-18-22852",
    doi = "10.1103/PhysRevC.98.014322",
    journal = "Phys. Rev. C",
    volume = "98",
    number = "1",
    pages = "014322",
    year = "2018"
}

@article{Schiavilla:2006xx,
    author = "Schiavilla, R. and Wiringa, Robert B. and Pieper, Steven C. and Carlson, J.",
    title = "{Tensor Forces and the Ground-State Structure of Nuclei}",
    eprint = "nucl-th/0611037",
    archivePrefix = "arXiv",
    reportNumber = "JLAB-THY-06-562",
    doi = "10.1103/PhysRevLett.98.132501",
    journal = "Phys. Rev. Lett.",
    volume = "98",
    pages = "132501",
    year = "2007"
}

@article{CiofidegliAtti:1995qe,
    author = "Ciofi degli Atti, Claudio and Simula, S.",
    title = "{Realistic model of the nucleon spectral function in few and many nucleon systems}",
    eprint = "nucl-th/9507024",
    archivePrefix = "arXiv",
    doi = "10.1103/PhysRevC.53.1689",
    journal = "Phys. Rev. C",
    volume = "53",
    pages = "1689",
    year = "1996"
}

@article{Weiss:2016obx,
    author = "Weiss, R. and Cruz-Torres, R. and Barnea, N. and Piasetzky, E. and Hen, O.",
    title = "{The nuclear contacts and short range correlations in nuclei}",
    eprint = "1612.00923",
    archivePrefix = "arXiv",
    primaryClass = "nucl-th",
    doi = "10.1016/j.physletb.2018.01.061",
    journal = "Phys. Lett. B",
    volume = "780",
    pages = "211--215",
    year = "2018"
}

@article{Frankfurt:2008zv,
    author = "Frankfurt, Leonid and Sargsian, Misak and Strikman, Mark",
    title = "{Recent observation of short range nucleon correlations in nuclei and their implications for the structure of nuclei and neutron stars}",
    eprint = "0806.4412",
    archivePrefix = "arXiv",
    primaryClass = "nucl-th",
    doi = "10.1142/S0217751X08041207",
    journal = "Int. J. Mod. Phys. A",
    volume = "23",
    pages = "2991--3055",
    year = "2008"
}

@article{Benhar:1994hw,
    author = "Benhar, O. and Fabrocini, A. and Fantoni, S. and Sick, I.",
    title = "{Spectral function of finite nuclei and scattering of GeV electrons}",
    doi = "10.1016/0375-9474(94)90920-2",
    journal = "Nucl. Phys. A",
    volume = "579",
    pages = "493--517",
    year = "1994"
}

@article{Wei:2019xdw,
    author = "Wei, Gao-Feng and Zhi, Qi-Jun and Cao, Xin-Wei and Long, Zheng-Wen",
    title = "{Examination of an isospin-dependent single-nucleon momentum distribution for isospin-asymmetric nuclear matter in heavy-ion collisions}",
    eprint = "1901.10391",
    archivePrefix = "arXiv",
    primaryClass = "nucl-th",
    doi = "10.1007/s41365-020-00779-6",
    journal = "Nucl. Sci. Tech.",
    volume = "31",
    number = "7",
    pages = "71",
    year = "2020"
}

@article{Cai:2025txx,
    author = "Cai, Bao-Jun and Li, Bao-An and Ma, Yu-Gang",
    title = "{Nucleon Short-Range Correlations and High-Momentum Dynamics: Implications on the Equation of State of Dense Matter}",
    eprint = "2512.04206",
    archivePrefix = "arXiv",
    primaryClass = "nucl-th",
    doi = "10.1140/epjs/s11734-026-02227-9",
    journal = "Eur. Phys. J. Spec. Top.",
    year = "2026"
}

@article{Broniowski:2010jd,
    author = "Broniowski, Wojciech and Rybczynski, Maciej",
    title = "{Two-body nucleon-nucleon correlations in Glauber models of relativistic heavy-ion collisions}",
    eprint = "1003.1088",
    archivePrefix = "arXiv",
    primaryClass = "nucl-th",
    doi = "10.1103/PhysRevC.81.064909",
    journal = "Phys. Rev. C",
    volume = "81",
    pages = "064909",
    year = "2010"
}

@article{Alvioli:2011sk,
    author = "Alvioli, M. and Holopainen, H. and Eskola, K. J. and Strikman, M.",
    title = "{Initial state anisotropies and their uncertainties in ultrarelativistic heavy-ion collisions from the Monte Carlo Glauber model}",
    eprint = "1112.5306",
    archivePrefix = "arXiv",
    primaryClass = "hep-ph",
    doi = "10.1103/PhysRevC.85.034902",
    journal = "Phys. Rev. C",
    volume = "85",
    pages = "034902",
    year = "2012"
}

@article{Buss:2011mx,
    author = "Buss, O. and Gaitanos, T. and Gallmeister, K. and van Hees, H. and Kaskulov, M. and Lalakulich, O. and Larionov, A. B. and Leitner, T. and Weil, J. and Mosel, U.",
    title = "{Transport-theoretical Description of Nuclear Reactions}",
    eprint = "1106.1344",
    archivePrefix = "arXiv",
    primaryClass = "hep-ph",
    doi = "10.1016/j.physrep.2011.12.001",
    journal = "Phys. Rept.",
    volume = "512",
    pages = "1--124",
    year = "2012"
}

@article{Rios:2009gb,
    author = "Rios, Arnau and Polls, Artur and Dickhoff, W. H.",
    title = "{Depletion of the nuclear Fermi sea}",
    eprint = "0904.2183",
    archivePrefix = "arXiv",
    primaryClass = "nucl-th",
    doi = "10.1103/PhysRevC.79.064308",
    journal = "Phys. Rev. C",
    volume = "79",
    pages = "064308",
    year = "2009"
}

@article{Yong:2015gma,
    author = "Yong, Gao-Chan",
    title = "{Constraining nucleon high momentum in nuclei}",
    eprint = "1503.08523",
    archivePrefix = "arXiv",
    primaryClass = "nucl-th",
    doi = "10.1016/j.physletb.2016.12.013",
    journal = "Phys. Lett. B",
    volume = "765",
    pages = "104--108",
    year = "2017"
}

@article{Yong:2021nwn,
    author = "Yong, Gao-Chan",
    title = "{Symmetry energy extracted from the S{\ensuremath{\pi}}RIT pion data in Sn+Sn systems}",
    eprint = "2104.12098",
    archivePrefix = "arXiv",
    primaryClass = "nucl-th",
    doi = "10.1103/PhysRevC.104.014613",
    journal = "Phys. Rev. C",
    volume = "104",
    number = "1",
    pages = "014613",
    year = "2021"
}

@misc{srcdataset,
  author  = {Huang, Y.-J.},
  title   = {Effects of Initial-State Nucleon Short-Range Correlations on Pion Production and Proton Flow in Au+Au Collisions [dataset]},
  year      = {2026},
  howpublished = {figshare dataset},
  doi     = {10.6084/m9.figshare.35076097}
}

@article{Zhang2025AbInitioOO,
  author = {Zhang, Chunjian and Chen, Jinhui and Giacalone, Giuliano
            and Huang, Shengli and Jia, Jiangyong and Ma, Yu-Gang},
  title = {Ab-initio nucleon-nucleon correlations and their impact
           on high energy {$^{16}$O+$^{16}$O} collisions},
  journal = {Phys. Lett. B},
  volume = {862},
  pages = {139322},
  year = {2025},
  doi = {10.1016/j.physletb.2025.139322},
  eprint = {2404.08385},
  archivePrefix = {arXiv},
  primaryClass = {nucl-th}
}

@article{Arrington:2011xs,
    author = "Arrington, J. and Higinbotham, D. W. and Rosner, G. and Sargsian, M.",
    title = "{Hard probes of short-range nucleon-nucleon correlations}",
    eprint = "1104.1196",
    archivePrefix = "arXiv",
    primaryClass = "nucl-ex",
    reportNumber = "PHY-12946-ME-2011, JLAB-PHY-11-1329",
    doi = "10.1016/j.ppnp.2012.04.002",
    journal = "Prog. Part. Nucl. Phys.",
    volume = "67",
    pages = "898--938",
    year = "2012"
}

\end{document}